# Charge Transport in and Electroluminescence from sp$^3$-Functionalized Carbon Nanotube Networks

*Nicolas F. Zorn[1], Felix J. Berger[1], and Jana Zaumseil[1*]*

[1] Institute for Physical Chemistry and Centre for Advanced Materials, Universität Heidelberg,

D-69120 Heidelberg, Germany

Corresponding Author

*E-mail: zaumseil@uni-heidelberg.de

ABSTRACT

The controlled covalent functionalization of semiconducting single-walled carbon nanotubes (SWCNTs) with luminescent $sp^3$ defects leads to additional narrow and tunable photoluminescence features in the near-infrared and even enables single-photon emission at room temperature, thus strongly expanding their application potential. However, the successful integration of $sp^3$-functionalized SWCNTs in optoelectronic devices with efficient defect state electroluminescence not only requires control over their emission properties but also a detailed understanding of the impact of functionalization on their electrical performance, especially in dense networks. Here, we demonstrate ambipolar, light-emitting field-effect transistors based on networks of pristine and functionalized polymer-sorted (6,5) SWCNTs. We investigate the influence of $sp^3$ defects on charge transport by employing electroluminescence and (charge-modulated) photoluminescence spectroscopy combined with temperature-dependent current-voltage measurements. We find that $sp^3$-functionalized SWCNTs actively participate in charge transport within the network as mobile carriers efficiently sample the $sp^3$ defects, which act as shallow trap states. While both hole and electron mobilities decrease with increasing degree of functionalization, the transistors remain fully operational, showing electroluminescence from the defect states that can be tuned by the defect density.

Their high ambipolar charge carrier mobilities combined with narrow emission bands in the near-infrared (nIR) make single-walled carbon nanotubes (SWCNTs) a promising material for applications in electrically driven light sources.[1-4] Due to advances in sorting techniques such as selective wrapping with conjugated polymers,[5-7] it is now possible to prepare large amounts of chirality-pure SWCNT dispersions in organic solvents that are suitable for reproducible solution-processing of thin-films.[8-11] However, the progress in optoelectronic applications has been limited by the low luminescence efficiencies of SWCNT films (photoluminescence quantum yield ~0.1 %) and thin-film devices (external quantum efficiency < 0.01 %).[2, 3] These low emission efficiencies result from diffusion-limited exciton quenching, *e.g.*, at lattice defects or nanotube-nanotube junctions,[12] self-absorption due to a small Stokes shift,[7] the presence of low-lying dark excitonic states,[13] and Auger-type quenching by charges in devices.[14-16] Likewise, defect-induced exciton-quenching has been shown to limit exciton transport and harvesting efficiency in photovoltaic cells based on bilayers of fullerenes and semiconducting SWCNT networks.[17]

Recently, the controlled functionalization of SWCNTs with luminescent $sp^3$ defects has emerged as a powerful approach to enhance their optical properties.[18-20] These quantum defects can be created through covalent binding of functional groups such as oxygen,[21] alkyl,[22] or aryl moieties[23] and act as efficient exciton traps. Optical trap depths are on the order of ~100 – 300 meV. They are partially determined by the chemical nature of the attached groups[22] but more importantly by the defect binding configuration on the nanotube lattice.[24, 25] Through exciton localization, these $sp^3$ defects prevent diffusion to quenching sites and give rise to red-shifted photoluminescence (PL) with long lifetimes.[26] They even enable high-purity single-photon emission at room temperature.[27] Combined with the chirality-dependent emission wavelengths of nanotubes, $sp^3$ functionalization allows the emission to be spectrally tuned across the nIR for potential applications such as electrically pumped single-photon sources at telecommunication wavelengths.[28, 29]

While the spectroscopic properties of sp$^3$-functionalized SWCNTs have been extensively investigated, only very few studies were directed at the impact of functionalization on charge transport. Wilson *et al.* monitored the formation of sp$^3$ defects by measuring the resistance of electrically contacted, individual SWCNTs exposed to an aqueous diazonium salt solution.[30] In good agreement with quantum theoretical calculations,[31] an average resistance increase of ~6 kΩ was attributed to the introduction of a single defect. Similarly, Bouilly *et al.* observed a conductance drop in single-nanotube transistors between ~20 % and two orders of magnitude depending on the number of sp$^3$ defects created by diazonium chemistry in patterned nanowells.[32] Single-nanotube transistors with isolated sp$^3$ defects were further used as sensors by covalently attaching groups with specific binding motifs (*e.g.*, nucleic acid strands for DNA detection) and monitoring variations in the gating behaviour.[32, 33]

Despite these advances, a comprehensive picture of charge transport in sp$^3$-functionalized SWCNTs is still missing. This is mainly due to a lack of information on the investigated nanotube species, *i.e.*, their chirality or even their electronic type, in previous studies. Furthermore, in the absence of spectroscopic data, it was unclear whether or not the introduced defects were indeed luminescent exciton traps. Hence, the interaction of sp$^3$ defects and defect-localized excitons with charge carriers in active devices remains ambiguous. To the best of our knowledge, the impact of sp$^3$ functionalization on charge transport in semiconducting SWCNT networks that show excellent performances in field-effect transistors (FETs) has not been explored yet. A fundamental understanding of both their spectroscopic and electrical properties is, however, crucial to realize optoelectronic applications with sp$^3$-functionalized SWCNT networks.

Here, we demonstrate ambipolar and light-emitting FETs based on networks of monochiral and sp$^3$-functionalized (6,5) SWCNTs that show both high charge carrier mobilities and electroluminescence (EL) from the defect states. By employing EL, PL and charge-modulated PL spectroscopy in combination with temperature-dependent current-voltage measurements, we explore the interactions of sp$^3$ defects with charge carriers and their impact on charge transport through the network. We find that sp$^3$ defects act as shallow charge traps, from which carriers detrap easily at room temperature. Thus, sp$^3$-functionalized SWCNTs actively participate in charge transport through the network and contribute to the EL, allowing for the defect emission to be harnessed in nIR light-emitting devices.

RESULTS AND DISCUSSION

**sp$^3$ Functionalization of (6,5) SWCNTs.** We investigate highly purified semiconducting (6,5) SWCNTs as a model system to study the impact of sp$^3$ functionalization on charge transport through carbon nanotube networks. (6,5) SWCNT dispersions in toluene were obtained *via* shear force mixing and polymer-wrapping with the polyfluorene-bipyridine copolymer PFO-BPy (see **Figure 1a** and Methods).[7] Raman and absorption spectra (Supporting Information, **Figure S1**) confirmed the absence of metallic and other minority SWCNTs in the dispersion. The controlled introduction of luminescent sp$^3$ defects to polymer-wrapped nanotubes in organic solvents was accomplished by a recently established reaction protocol (**Figure 1a**).[34] This method employs the phase-transfer agent 18-crown-6 to solubilize a pre-formed diazonium salt (here, 4-bromobenzenediazonium tetrafluoroborate) in a (6,5) SWCNT dispersion in toluene/acetonitrile (80:20 vol.-%, for details see Methods). After reacting for 16 hours at room temperature and in the dark, nanotubes were collected *via* filtration and thoroughly washed to remove unreacted diazonium salt and reaction byproducts. Redispersion of the filter cake in a small volume of fresh toluene by bath sonication yielded concentrated dispersions suitable for spin-coating of dense SWCNT networks. Note that filtration and redispersion were similarly carried out before device processing for unfunctionalized ("pristine") (6,5) SWCNTs to remove most of the excess wrapping polymer.

Upon covalent functionalization, a new emission band ($E_{11}$*) appeared in the PL spectra of the SWCNT dispersions ($E_{22}$ excitation at 575 nm) in addition to the $E_{11}$ transition, as shown in **Figure 1b**. This additional peak originates from the radiative decay of excitons that are localized at symmetry-breaking sp$^3$ defects in the nanotube lattice.[23] For the bromoaryl defects employed here, the main defect PL peak occurred at ~1166 nm corresponding to a red-shift by ~174 meV from the mobile $E_{11}$ exciton (~1002 nm). The sp$^3$ defect density on the (6,5) SWCNTs and thus intensity of the $E_{11}$* emission was controlled by variation of the diazonium salt concentration in the reaction mixture (see **Figure 1b**). Higher concentrations resulted in higher $E_{11}$*/$E_{11}$ PL ratios. We estimate that the defect densities studied here range from approximately 5 to 30 defects per µm. For the highest degree of functionalization, an additional emission band ($E_{11}$*$^-$) - even further red-shifted than the dominant $E_{11}$* transition - was observed in agreement with previous results.[34] This feature is attributed to a different defect binding configuration on the chiral (6,5) SWCNT lattice, resulting in deeper exciton traps as indicated by quantum mechanical calculations.[35, 36]

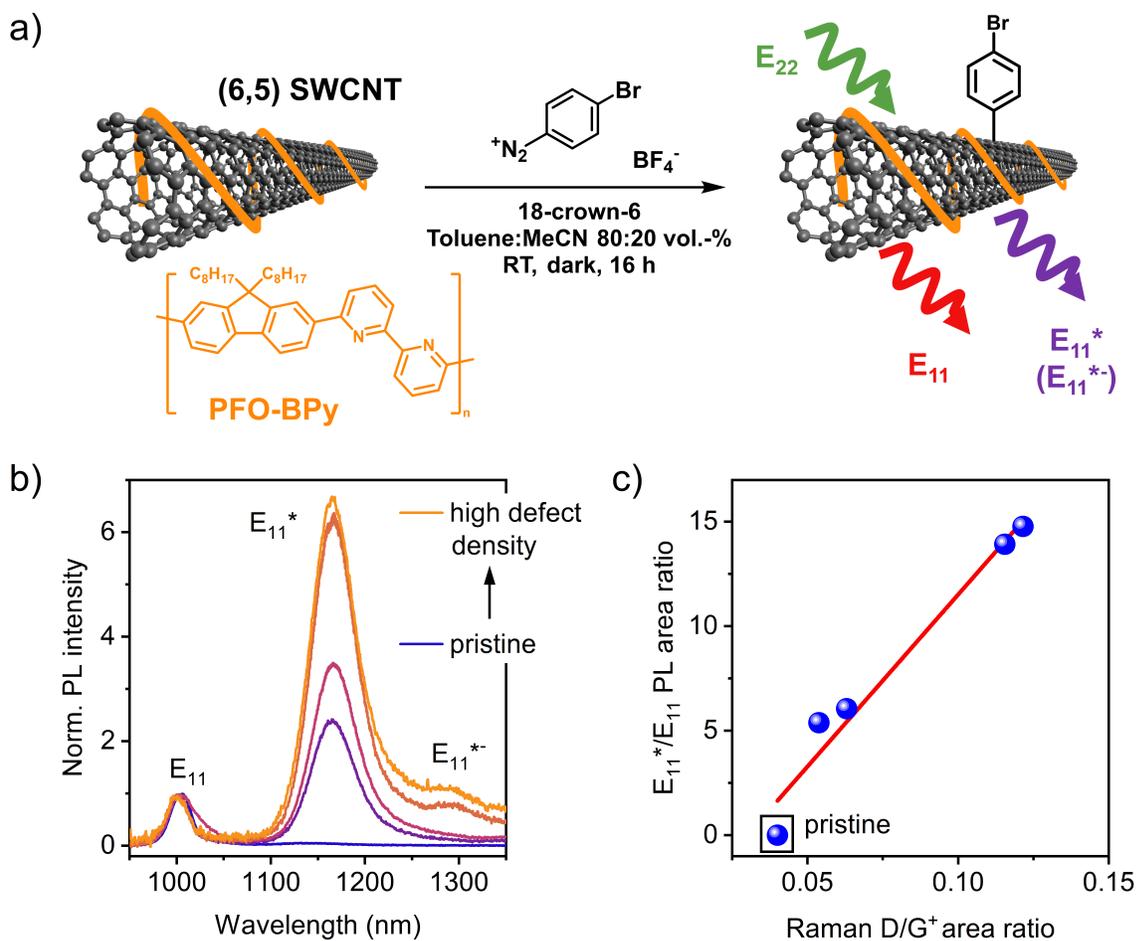

**Figure 1. a)** Reaction scheme for the controlled sp$^3$ functionalization of PFO-BPy-wrapped (6,5) SWCNTs with 4-bromobenzenediazonium tetrafluoroborate and schematic illustration of the PL properties of functionalized SWCNTs. After optical E$_{22}$ excitation, red-shifted E$_{11}$* and E$_{11}$*$^-$ emission from defect-localized excitons is observed in addition to E$_{11}$ emission. **b)** Normalized PL spectra of sp$^3$-functionalized (6,5) SWCNT dispersions with different defect densities under pulsed excitation at the E$_{22}$ transition (575 nm, ~ 0.02 mJ cm$^{-2}$). **c)** Correlation between integrated PL ratio and Raman D/G$^+$ ratio and linear fit to the data.

The controlled variation of the defect densities was further corroborated by resonant Raman spectroscopy. An increasing ratio of the defect-related D mode to $G^+$ mode Raman signal areas (Raman D/$G^+$ area ratio) was observed with rising diazonium salt concentration and can be used as a metric for the number of $sp^3$ defects in the nanotube lattice (Supporting Information, **Figure S2**). For the defect density range explored here, the integrated $E_{11}*/E_{11}$ PL ratio also correlated well with the Raman D/$G^+$ ratio (**Figure 1c**). In addition, the recorded absorption spectra clearly showed the emergence of a $E_{11}*$ defect state absorption band at ~1156 nm (Supporting Information, **Figure S3**) that increased with defect density. For further characterization, the Raman D/$G^+$ ratio as well as the integrated $E_{11}*/E_{11}$ ratio of the absorption and PL spectra were used as metrics for the defect concentration.[34]

**Field-effect transistors with $sp^3$-functionalized SWCNT networks.** To investigate the properties of $sp^3$-functionalized SWCNTs in active optoelectronic devices and to determine the impact of functionalization on charge transport (*e.g.*, scattering or trapping of charge carriers at the defects, as schematically illustrated in **Figure 2a**), we fabricated bottom-contact, top-gate field-effect transistors (FETs, see **Figure 2b** for schematic device layout) with interdigitated source/drain electrodes (channel length, $L$ = 20 μm and channel width, $W$ = 10 mm), dense spin-coated layers of pristine and $sp^3$-functionalized SWCNTs (see **Figure 2c** and Supporting Information, **Figure S4** for atomic force microscopy (AFM) images), a bilayer PMMA/HfO$_x$ dielectric, and a silver top-gate electrode (see Methods section for details). All SWCNT networks were annealed in dry nitrogen at 150 °C to remove residual water and oxygen while avoiding detachment of covalently bound aryl groups above ~200 °C as recently shown by Schirowski *et al.*[37]

**Figure 2d** shows the transfer characteristics in the linear regime (source-drain voltage $V_{ds}$ = -100 mV) for a pristine SWCNT network FET and a device with an sp$^3$-functionalized nanotube network with high defect density. Both clearly show ambipolar charge transport (*i.e.*, electron and hole conduction) with on/off current ratios on the order of 10$^5$ – 10$^6$ and low gate leakage currents (< 1 nA). A significant decrease in both electron and hole currents and thus carrier mobility is evident for the highly functionalized SWCNT network FET in comparison to the pristine reference. Corresponding output characteristics as well as transfer curves for network FETs with all of the different defect densities are shown in the Supporting Information, **Figures S5, S6**. The slight current hysteresis in all transfer characteristics probably results from the relatively low annealing temperature and hence electron trapping by residual moisture,[38] which is also reflected in the lower electron currents compared to the hole currents even for pristine SWCNTs.

To evaluate the impact of sp$^3$ functionalization on the charge transport in these SWCNT networks, we first compare the maximum carrier mobilities in the linear regime. As shown in **Figure 2e**, the linear electron and hole mobilities gradually decrease with increasing defect density, here represented by the Raman D/G$^+$ area ratio. Linear hole mobilities reached 5.0 ± 0.2 cm$^2$ V$^{-1}$ s$^{-1}$ for pristine and 1.7 ± 0.1 cm$^2$ V$^{-1}$ s$^{-1}$ for highly functionalized networks, whereas electron mobilities decreased from 1.2 ± 0.2 cm$^2$ V$^{-1}$ s$^{-1}$ (pristine) to 0.4 ± 0.04 cm$^2$ V$^{-1}$ s$^{-1}$ (highest defect density). Note that for functionalized networks higher gate voltages had to be applied to reach the maximum transconductance and thus charge carrier mobility (see Supporting Information, **Figure S5b**). **Figure 2f** shows the decrease of hole and electron mobilities normalized to the pristine reference. Both drop to roughly one third of the initial value for the highest defect concentration with a Raman

D/G$^+$ area ratio of 0.12 and with an E$_{11}$*/E$_{11}$ PL ratio of ~15 in dispersion (see Supporting Information, **Figure S7** for absolute and normalized mobilities *vs.* various defect metrics).

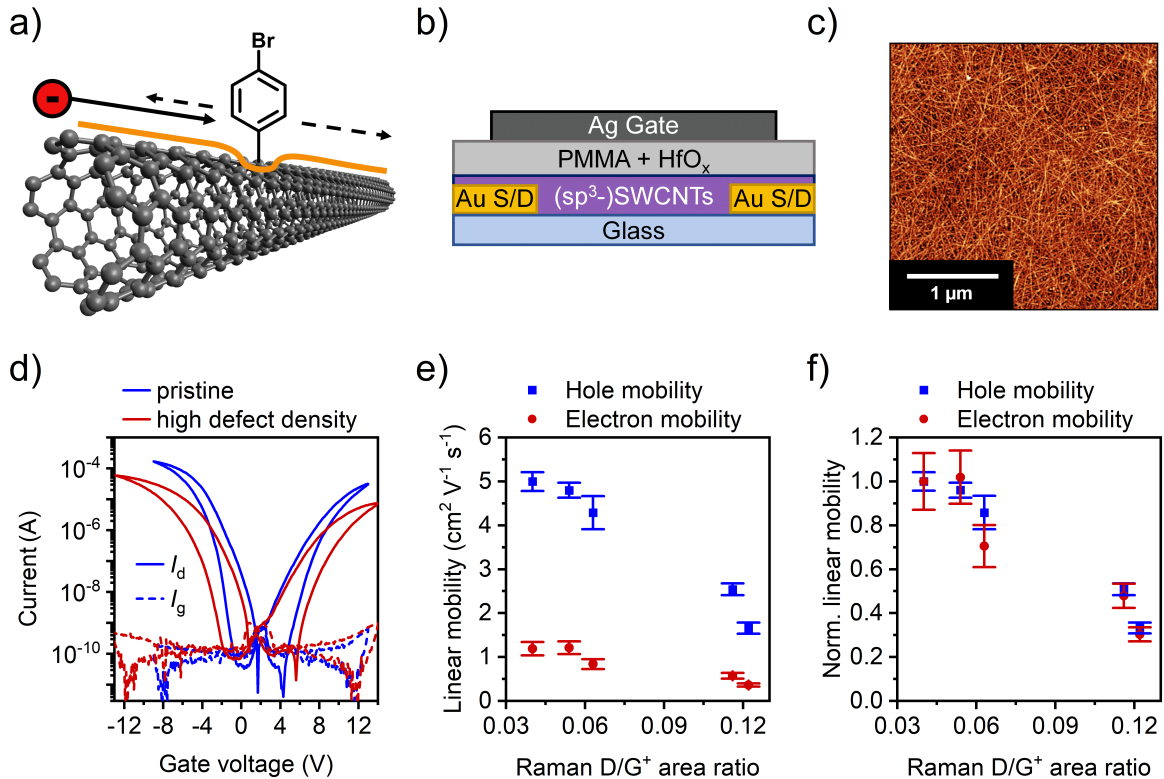

**Figure 2. a)** Schematic illustration of charge transport through sp$^3$-functionalized SWCNTs. Defects may act as scattering sites or trap states for charge carriers. **b)** Schematic device architecture of bottom-contact, top-gate SWCNT network FETs. **c)** Atomic force micrograph (scalebar, 1 µm) of a representative, dense network of sp$^3$-functionalized (6,5) SWCNTs. **d)** Ambipolar transfer characteristics (source-drain voltage $V_{ds}$ = -100 mV, $L$ = 20 µm, $W$ = 10 mm) of FETs with networks of pristine (blue) and sp$^3$-functionalized (6,5) SWCNTs (red). Solid lines are drain currents, $I_d$, dashed lines are gate leakage currents, $I_g$. **e)** Absolute and **f)** normalized linear charge carrier mobilities (holes, blue squares; electrons, red circles) of pristine and sp$^3$-functionalized (6,5) SWCNT FETs *vs.* Raman D/G$^+$ area ratio.

Frequency-dependent capacitance measurements further showed a decrease in the FET cut-off frequency with increasing degree of functionalization, mainly reflecting the lower carrier mobilities of the network (Supporting Information, **Figure S8**). The static capacitances of all network FETs, as extracted from capacitance-voltage sweeps, were very similar (Supporting Information, **Table S1**), confirming similar network densities that were also well above the limit for mobility saturation.[39] Hence, any observed differences in electrical performance can be attributed directly to the different $sp^3$ defect densities.

As shown above, even at the highest level of $sp^3$ functionalization studied here, the SWCNT network FETs remain fully operational and still exhibit surprisingly high ambipolar carrier mobilities. In previous reports, the introduction of a single $sp^3$ defect on an individual semiconducting nanotube already resulted in a conductance drop of ~20 %.[30, 32] However, in single-nanotube devices, charge carriers are forced to pass $sp^3$ defects, whereas this is not necessarily the case in random nanotube networks with multiple possible conduction pathways. This difference raises fundamental questions about the charge transport mechanisms in functionalized SWCNT networks: First, what is the nature of $sp^3$ defects with respect to charge transport along one nanotube (scattering site, shallow or deep trap state)? And second, do $sp^3$-functionalized SWCNTs still actively participate in charge transport or are nanotube segments with defects bypassed in a dense network? The latter would result in lower overall mobilities simply due to fewer available current pathways but would also reduce possible electroluminescence from defect sites. These scenarios directly affect the application potential of $sp^3$-functionalized SWCNT networks for electroluminescent devices. They will be addressed in the following by comparing electroluminescence measurements, static gate voltage-dependent photoluminescence, charge modulation photoluminescence spectroscopy, and temperature-dependent transport measurements of pristine and functionalized (6,5) SWCNT network FETs.

**Electroluminescence from sp$^3$ defect sites.** To investigate the charge transport pathways through networks of pristine and sp$^3$-functionalized (6,5) SWCNTs, we employed electroluminescence (EL) measurements in the near-infrared. The injection, transport and recombination of holes and electrons in the ambipolar regime leads to the generation of excitons and thus light emission,[40] as schematically illustrated in **Figure 3a**. EL is observed from a narrow recombination zone that is formed where the hole and electron accumulation layers meet in the channel. Since EL correlates directly with current density, its spatial and spectral resolution has been previously utilized to analyze charge transport through mixed-chirality SWCNT networks *via* the bandgap-dependent current shares,[15, 16] and to visualize patterned current pathways in photoswitchable SWCNT network FETs.[41] Here, we use the spectral signatures of sp$^3$ defects (E$_{11}$*) and mobile excitons (E$_{11}$) to study the contribution of functionalized nanotubes to charge transport in the networks.

First, we imaged the near-infrared EL from these FETs with a 2D-InGaAs camera (800 – 1600 nm) during a constant current sweep ($I_d$ = -100 µA) in the ambipolar regime. **Figure 3b** shows a homogeneous recombination and emission zone within the channel ($L$ = 20 µm) of a sp$^3$-functionalized SWCNT FET. This narrow emission zone (width 1 – 2 µm) extended without interruption along the entire channel width and could be arbitrarily positioned between source and drain electrodes by adjusting the gate voltage. Adding up a large number of such images for a complete gate voltage sweep yields a composite EL image (see **Figure 3c**) that reflects the current density distribution within the channel.[41, 42] No preferential transport paths were observed on the length scale of this diffraction-limited imaging method.

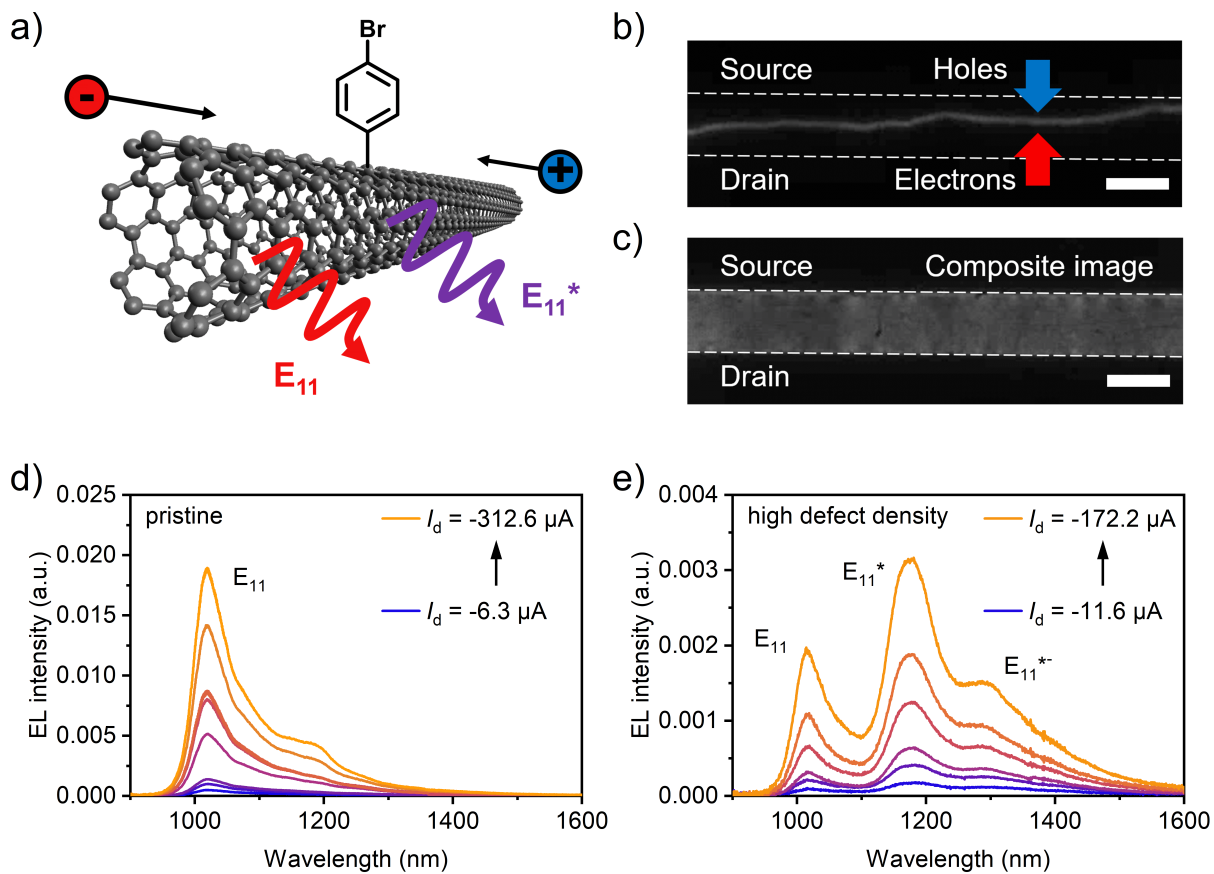

**Figure 3. a)** Schematic illustration of electroluminescence from sp$^3$-functionalized SWCNTs through ambipolar carrier recombination of thermalized holes and electrons. **b)** Near-infrared EL image of the channel ($L$ = 20 µm) of an sp$^3$-functionalized SWCNT network FET, showing a homogeneous recombination and emission zone when the device is biased in the ambipolar regime (drain current $I_d$ = -100 µA). **c)** Composite EL image for a full gate voltage sweep at a constant current ($I_d$ = -100 µA), showing homogeneous EL emission from the entire channel area (scalebars, 20 µm). **d, e)** Representative EL spectra of pristine and sp$^3$-functionalized (high defect density) SWCNT network FETs in the ambipolar regime for different drain currents.

We then resolved the EL spectrally with the emission zone always positioned in the center of the channel (as in **Figure 3b**) for different drain currents. Note that changing the drain current while keeping the position of the emission zone in the center of the channel requires changing both the gate voltage $V_\text{g}$ and the source-drain voltage $V_\text{ds}$. **Figure 3d** shows the EL spectra of a pristine SWCNT network transistor. In addition to the $E_{11}$ emission, the spectra exhibit a prominent tail toward longer wavelengths that can be attributed to typical SWCNT emission side bands.[43, 44] Trion emission at ~1180 nm, which is strong in electrolyte-gated SWCNT FETs,[45] is rather weak here, probably due to the significantly lower carrier densities. As expected, the emission intensity increased nearly linearly with the drain current. Maximum EL emission efficiencies of ~0.015 % were obtained for these pristine SWCNT networks.

For FETs with $sp^3$-functionalized SWCNTs, electrically induced emission from the $sp^3$ defects is clearly observed (see **Figure 3e**). As shown in the Supporting Information, **Figure S9**, the defect state EL is tunable in intensity and increases with the level of functionalization. At the highest defect density, about 80 % of emission comes from $sp^3$ defects. Although we observed a slight increase in EL emission efficiencies for low defect densities, they largely remained within the same range as the unfunctionalized SWCNT network devices and slightly decreased for the highest defect densities as expected. The spectral shape of the EL was also essentially independent of the position of the emission zone within the transistor channel, even at the electrodes. Normalizing the EL spectra to the $E_{11}$ exciton signal confirms that the defect emission is stable over 1 – 2 orders of magnitude in current density, which should be beneficial for applications (Supporting Information, **Figure S10**). Only a slight decrease in the $E_{11}*/E_{11}$ peak ratio was observed for increasing drain current (and applied gate voltage) for all functionalized networks.

Nevertheless, the ratio of defect-to-$E_{11}$ electroluminescence was significantly lower than the PL ratio in the original dispersion (see **Figure 1b**). One possible cause might be different excitation densities and thus state-filling of exciton traps, which is well-known from the PL spectra of $sp^3$-functionalized SWCNTs. The defect emission shows earlier saturation compared to the mobile $E_{11}$ excitons for increasing pump power.[34, 46] However, a plot of the integrated EL intensities *vs.* drain currents on a log-log scale yields slopes of ~1 for both the defect and $E_{11}$ emission peaks (Supporting Information, **Figure S11a**), which confirms that the data was collected in the linear excitation regime without significant state-filling. This notion is further supported by the relatively low exciton densities (for details and calculations see the Supporting Information) that are estimated for EL at the given current densities. Additionally, a PL spectrum acquired from the channel region under non-resonant, low-power continuous wave (cw) excitation displays very similar $E_{11}^*/E_{11}$ peak ratios (Supporting Information, **Figure S11b**). The lower defect emission intensity appears to be mainly associated with film formation.

Interestingly, emission from the even further red-shifted $E_{11}^{*-}$ defects, which originate from a different defect binding configuration, is more prominent in the EL spectra of highly functionalized SWCNT networks compared to the PL spectrum. The ultrafast exciton transfer within nanotube films[47] should affect both EL and PL spectra in a similar fashion. To address this question, an investigation of the emission properties of SWCNT networks with both $E_{11}^*$ and $E_{11}^{*-}$ defects at similar densities, as can be created by a recently reported base-promoted functionalization method,[25] would be useful but is beyond the scope of the current study.

Note that the $sp^3$ defect EL observed here originates from the recombination of thermalized electrons and holes and not from impact excitation as shown recently by Xu *et al.*[48] Our results demonstrate that $sp^3$-functionalized SWCNTs can be integrated in multi-functional devices such as light-emitting FETs, where they support ambipolar charge transport and exhibit electroluminescence. The observation of defect emission indicates that electron-hole recombination occurs near (within the exciton diffusion length) or at defect sites and hence charge transport must at least partially go through segments of nanotubes bearing sp$^3$ defects. The similarity of the EL and PL spectrum of functionalized SWCNT networks at low excitation densities suggests that there are no preferential transport paths or recombination sites.

**Static, gate-voltage dependent photoluminescence.** A more detailed analysis of photoluminescence in FETs should help to elucidate the interaction between excitons and charge carriers in sp$^3$-functionalized SWCNT networks and their impact on emission properties. We performed static, gate voltage-dependent PL measurements under non-resonant cw excitation (785 nm) in the linear excitation regime (**Figure 4** and Supporting Information, **Figure S12**) as well as under pulsed $E_{22}$ excitation (Supporting Information, **Figures S13, S14**). In the latter case, the relative $E_{11}^*$ and $E_{11}^{*-}$ peak intensities were considerably lower due to the much higher exciton densities (about two orders of magnitude, for details see the Supporting Information) leading to state filling.[34, 46]

All gate voltage-dependent PL spectra show a decrease of the emission intensities ($E_{11}$, $E_{11}^*$, $E_{11}^{*-}$) with increasing carrier density (holes or electrons) due to Auger-type quenching of excitons with charges (for PL spectra in hole accumulation see **Figure 4a**).[49] However, quenching of the defect PL is more efficient than that of the $E_{11}$ emission, as shown in the normalized spectra in **Figure 4b**. This trend is analogous to the EL spectra, where the $E_{11}^*/E_{11}$ ratio also decreases slightly for increasing current densities.

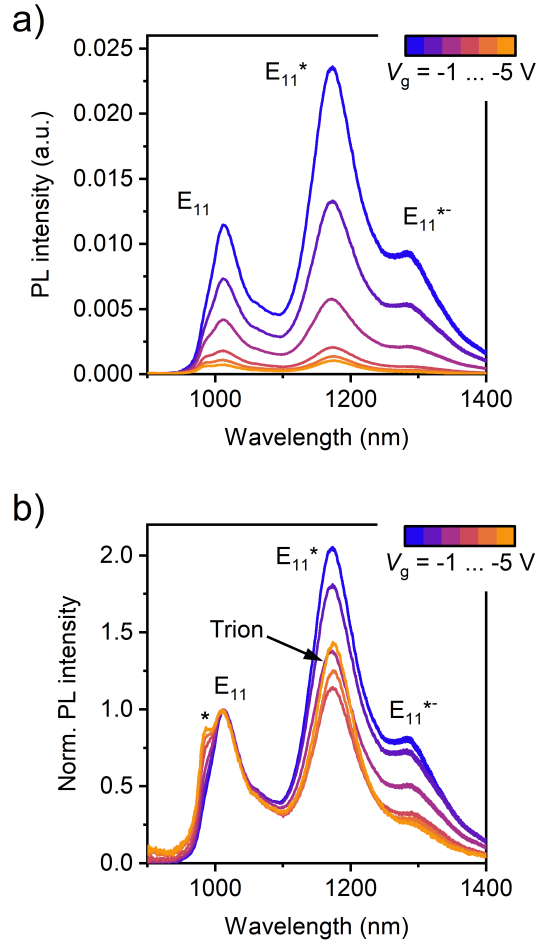

**Figure 4. a)** Static, gate voltage-dependent PL spectra (source-drain voltage $V_{ds}$ = -10 mV) of sp$^3$-functionalized SWCNT network transistors ($L$ = 20 µm) with high defect density in hole accumulation. Spectra were acquired from the middle of the channel under non-resonant continuous wave excitation (785 nm, ~320 W cm$^{-2}$). **b)** Normalized PL spectra show that $E_{11}$* and $E_{11}$*$^-$ defect PL is more efficiently quenched than the $E_{11}$ emission when gate voltages are applied. At high $V_g$, PL from positively charged trions can be observed at very similar wavelengths to the $E_{11}$* emission. Note that the peak at ~985 nm marked with an asterisk corresponds to the Raman 2D mode of (6,5) SWCNTs.

Previous studies by Shiraishi *et al.* on PL from electrochemically doped oxygen- and aryl-functionalized (6,5) SWCNT films revealed a similar effect and allowed them to determine redox potentials corresponding to the highest occupied molecular orbital (HOMO) and lowest unoccupied molecular orbital (LUMO) levels as well as the electrochemical bandgap of the defect sites.[50, 51] For bromoaryl defects (as used here) the electrochemical bandgap was ~53 meV smaller for $E_{11}$* compared to $E_{11}$ (HOMO level shifted by ~31 meV to higher energies, LUMO level shifted by ~22 meV to lower energies), suggesting that these $sp^3$ defects act as shallow charge carrier traps that would enable detrapping at room temperature (thermal energy ~25 meV). Charges that preferentially reside at the $sp^3$ defects may explain the predominant quenching of $E_{11}$* excitons and thus the decrease of the $E_{11}$*/$E_{11}$ ratio in PL and EL spectra with increasing hole or electron density.

Further evidence is provided by pump power-dependent PL measurements under cw excitation (Supporting Information, **Figures S15, S16**). The integrated $E_{11}$*/$E_{11}$ and $E_{11}$*$^-$/$E_{11}$ PL intensity ratios show that the saturation of defect emission occurs at higher excitation powers when a gate voltage ($V_g$ = -2 V, hole accumulation) is applied compared to the neutral state ($V_g$ = 0 V). Charge carriers residing in close proximity to the defect sites should lead to a shortening of localized-exciton lifetimes through non-radiative quenching. Hence, state-filling, which is commonly attributed to the long lifetime of $sp^3$ defect-localized excitons and fast exciton diffusion to defect sites,[46] is reduced and becomes significant only at higher excitation densities. Unfortunately, a quantitative evaluation of the PL spectra with regard to the defect-to-$E_{11}$ emission ratio depending on the carrier density (gate voltage) is hindered by the additional contribution of trion emission at very similar wavelengths (~1175 nm) to the $E_{11}$* peak (~1173 nm) (Supporting Information, **Figures S12-S14**).

Overall, the gate voltage-dependent PL data suggest a certain degree of charge localization at the sp$^3$ defects within the SWCNT networks. Due to the static nature of both EL and PL spectroscopy, the observed dependencies could result from either mobile or trapped charge carriers. To unambiguously investigate the interaction of sp$^3$ defects in (6,5) SWCNTs with mobile charge carriers, a dynamic measurement such as modulation spectroscopy is required.

**Charge modulation PL spectroscopy.** Charge modulation spectroscopy (CMS) is a dynamic and highly sensitive method that has been widely used to investigate charge transport in organic semiconductors.[52, 53] We recently applied this technique to SWCNT network FETs and introduced charge modulation PL (CMPL) spectroscopy as a complementary approach to probe the mobile charge carriers in SWCNT networks.[54] In CMPL, a sinusoidal voltage (peak-to-peak voltage $V_{pp}$) is applied to the gate electrode (which is at a constant offset voltage $V_{os}$) while the semiconducting layer is excited by a laser. The differential change in PL ($\Delta PL$) upon modulation of charge carrier density in the transistor channel is recorded with a lock-in detection scheme. The detected signal correlates exclusively with mobile charge carriers because charges in deep trap states cannot be modulated. A schematic of the setup is shown in the Supporting Information, **Figure S17**.

**Figure 5a** shows the CMPL spectra for a pristine (6,5) SWCNT network at different $V_{os}$ (modulation frequency $f$ = 363 Hz, $V_{pp}$ = 0.2 V). In agreement with previous results, the signal attributed to E$_{11}$ exciton quenching initially increases with $V_{os}$ due to the increase in capacitance and higher modulated carrier density.[54] At approximately -1.3 V, the $\Delta PL$ signal passes through a maximum and decreases again for higher $V_{os}$. Since the offset voltage $V_{os}$ corresponds to the static charge carrier density in the channel, on which the modulation is superimposed, this decrease can be attributed to static exciton quenching (see the static voltage-dependent PL spectra).

Consequently, the effect of charge modulation becomes less significant for high $V_{os}$ and because only the differential change is recorded in CMPL, the signal decreases. Spectra normalized to the $E_{11}$ peak are essentially identical (Supporting Information, **Figure S18a**).

Variation of the modulation frequency was used to probe the temporal response of charge carriers to the sinusoidal bias. Since the carrier mobilities in the investigated gate voltage range are still low (Supporting Information, **Figure S5b**), charges cannot fully follow high modulation frequencies and the signal intensity decreases with increasing frequency (Supporting Information, **Figure S18b**). The normalized spectra, however, are nearly identical (Supporting Information, **Figure S18c**), confirming that all contributions to the spectrum arise from the same physical effect, that is, quenching by mobile charges. Modulated PL spectra with good signal-to-noise ratios were obtained for frequencies as high as 10 kHz which is in good agreement with the cut-off frequencies observed in the capacitance measurements of the FETs (Supporting Information, **Figure S8**).

For $sp^3$-functionalized SWCNT networks, the voltage-dependent CMPL spectra closely resemble the static PL spectra, exhibiting signals corresponding to the modulation of $E_{11}$ and $sp^3$ defect emission. **Figure 5b** shows the CMPL spectra for a high defect density sample, where the peaks at ~1172 nm and ~1290 nm are assigned to the $E_{11}*$ and $E_{11}*^-$ transitions, respectively. Importantly, as CMPL features are attributed to the PL modulation by mobile carrier density, this observation corroborates that $sp^3$-functionalized SWCNT segments in a network are indeed sampled by mobile carriers. This notion is further supported by frequency-dependent CMPL measurements that showed modulation of defect PL emission even at frequencies of ~8 kHz (Supporting Information, **Figure S19**). The near-identical normalized spectra confirm that all spectral contributions have a common physical origin, that is, quenching by mobile charges.

Voltage- and frequency-dependent CMPL spectra of sp³-functionalized SWCNT network FETs with a low defect density also follow these trends (Supporting Information, **Figure S20**).

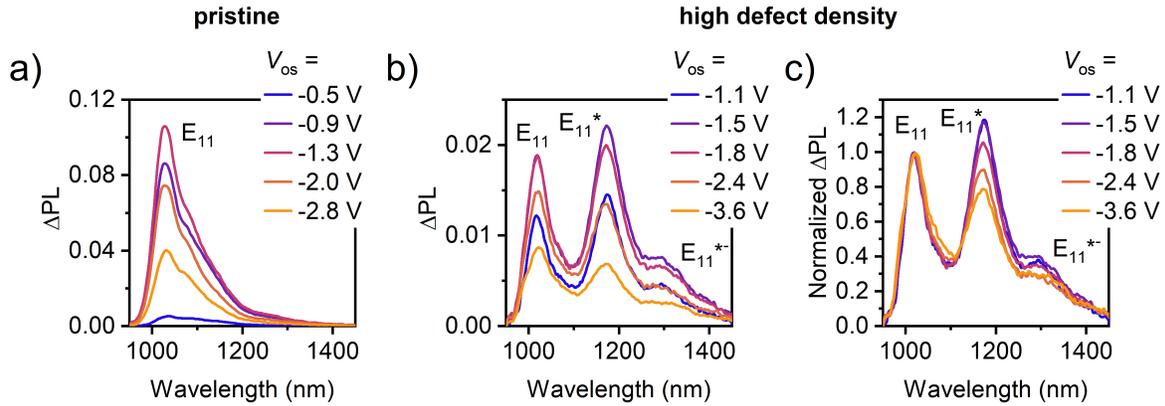

**Figure 5. a)** Voltage-dependent CMPL spectra of pristine (6,5) SWCNT network, showing an initial increase and subsequent decrease of the $E_{11}$ $\Delta PL$ signal with $V_{os}$. **b)** Voltage-dependent CMPL spectra of a sp³-functionalized SWCNT network with high defect density, showing PL modulation of mobile ($E_{11}$) as well as defect-localized ($E_{11}^*$, $E_{11}^{*-}$) excitons. **c)** CMPL spectra normalized to the $\Delta PL$ signal of $E_{11}$. All spectra were acquired from the middle of the channel ($L$ = 20 µm) at a modulation frequency $f$ = 363 Hz and $V_{pp}$ = 0.2 V.

Normalized CMPL spectra of FETs with functionalized (6,5) SWCNTs (**Figure 5c**) show a decrease in $E_{11}^*/E_{11}$ signal intensity with increasing charge carrier density (*i.e.*, $V_{os}$). This decrease is similar to the static voltage-dependent PL spectra (Supporting Information, **Figure S12**) and also suggests that the sp³ defects are lower-lying electronic states that participate in the charge transport through the networks. With increasing static carrier density, these shallow trap states are filled and thus cannot be fully modulated anymore.

At first glance, this picture seems to contradict the frequency-dependent data, which suggest that charge carriers are highly mobile on SWCNT segments with sp$^3$ defects under the measurement conditions. No CMPL response would be expected for carriers in deep trap states or at least some differences should appear in the temporal response at higher modulation frequencies. However, modulation spectroscopy is able to probe shallow traps as long as they can be filled and emptied on the timescale of charge modulation.[53] Short detrapping times at room temperature could thus explain this observation. Analogous to He *et al.* who calculated the thermal detrapping rates for excitons localized at sp$^3$ defects,[55] we estimate the thermal detrapping rate constant ($k$) for charge carriers using the Arrhenius equation $k = A \cdot \exp\left(-\frac{E_a}{k_\mathrm{B}T}\right)$. Based on transition state theory, the preexponential factor can be written as $A = k_\mathrm{B}T/h$ (with $k_\mathrm{B}$ – Boltzmann constant, $T$ – temperature, $h$ – Planck constant). Assuming a trap depth of ~25 meV, we find detrapping times on the order of ~1 ps at 300 K, which is orders of magnitude faster than the voltage modulation in CMPL (~10$^3$–10$^4$ Hz). Even assuming the optical trap depths of ~175 meV ($E_{11}$*) and ~255 meV ($E_{11}$*$^-$) still results in estimated detrapping times of ~100 ps to few ns. Lowering the temperature could lead to slower detrapping and might eventually result in changes in the modulation spectra, as shown for organic semiconductors in low-temperature CMS experiments.[53, 56] However, this experiment was not possible in our current CMPL setup.

**Temperature-dependent charge transport.** The data discussed so far have shown that sp$^3$ defects lead to red-shifted electroluminescence from SWCNT networks but also impede charge transport to some degree. Modulation spectroscopy at room temperature indicates that the defects are sampled by mobile carriers while acting as shallow traps, from which charges can detrap quickly at room temperature. The presence of such shallow trap states should be reflected in changes in the subthreshold regime of the linear transfer characteristics of network FETs even at ambient temperatures. Indeed, as shown in the Supporting Information, **Figure S21**, the subthreshold slope at 300 K was lower for the sp$^3$-functionalized SWCNT network transistors compared to the pristine (6,5) SWCNT FETs. We calculated the nominal trap densities from the subthreshold swing according to Kalb *et al.*[57] (see Supporting Information, **Table S2**). They are on the order of $10^{12}$ cm$^{-2}$ eV$^{-1}$ for the pristine (6,5) SWCNT networks as shown previously.[58] Upon functionalization, trap densities for electrons and holes increased by 50 – 60 %, indicating a noticeable but still moderate impact of the introduced sp$^3$ defects.

Temperature-dependent current-voltage measurements are a common tool to investigate charge transport in semiconductors in general and in semiconducting carbon nanotube networks in particular.[59-61] Shallow traps should have a larger effect at lower temperatures when detrapping is slowed down. For this purpose, FETs including two additional voltage probes in the channel were fabricated from the same SWCNT dispersions as discussed above to perform gated four-point-probe transport measurements (see Supporting Information, **Figure S22** for device architecture and measurement principle). This technique allows us to separate the temperature-dependent contact resistance from the channel resistance and thus to determine actual carrier mobilities in the nanotube network. This correction for contact resistance leads to a noticeable increase of the apparent network mobilities by 20 – 90 %. Hence, all of the following carrier mobilities are contact resistance-corrected.[61, 62]

To assess the temperature dependence of the network mobilities, transfer characteristics in the linear regime ($V_{ds}$ = -100 mV) were recorded from 300 K down to 25 K (see **Figure 6a** with only every other transfer curve shown) for a pristine (6,5) SWCNT network and for devices with low and high sp$^3$ defect densities. Note that different gate voltage ranges were used for unfunctionalized and functionalized samples to reduce hysteresis while still covering the full accumulation range. The on-currents monotonously increased with temperature, corroborating thermally activated charge transport in all networks as reported before.[61, 63] Due to shifts of the onset voltages $V_{on}$ with temperature, the contact resistance-corrected mobilities were extracted at a fixed gate voltage overdrive ($V_g$ - $V_{on}$) of ±6 V for electrons and holes, respectively, and thus at similar carrier concentrations (~4·10$^{12}$ cm$^{-2}$).

Temperature-dependent mobilities normalized to the respective value at 300 K are shown in **Figure 6b** for holes and electrons (for absolute values see Supporting Information, **Figure S23**). The logarithmic plots *vs.* 1/$T$ show pronounced differences between pristine, low defect density and high defect density SWCNT networks in the low-temperature range, but a very similar behavior at higher temperatures. These differences are also observed when comparing the maximum mobilities.

The variable range hopping (VRH) model, which was developed to describe charge transport by hopping in disordered semiconductors,[64] clearly does not fit to the data especially for low temperatures. The fluctuation-induced tunneling (FIT) model would give better fits and is more commonly used.[65] However, the corresponding fit results depend strongly on the starting values and are rather difficult to compare. In any case, both models associate the temperature dependence of the carrier mobilities only with thermally activated carrier hopping across nanotube-nanotube junctions and neglect the contribution of intra-nanotube transport, which is proportional to 1/$T$ for pristine SWCNTs.[66] We presume that the nanotube-nanotube junctions are unlikely to be directly affected by sp$^3$ defects at the low densities considered here with fewer than 3 defects per 100 nm (see above). If this assumption is correct, the observed differences between the absolute network mobility values and their temperature dependence must be caused by changes in the intra-nanotube conductance, further supporting the idea that charge transport in SWCNT networks is not solely limited by the junctions but represents a superposition of intra- and inter-nanotube contributions.[63] Note that the impact of sp$^3$ defects on nanotube bundling and electrical contact within bundles remains unclear. However, due to the small size of the bromoaryl moieties compared to the fluorene units with the octyl sidechains of the PFO-BPy wrapping polymer and the low defect densities, a significant effect seems unlikely.

Even without an analytical model combining intra- and inter-nanotube contributions, we can compare the relative differences in temperature dependence of the mobility data in **Figure 6b**. At higher temperatures, thermally activated hopping across the nanotube-nanotube junctions is the dominating process for carriers moving through the networks. Assuming that the junctions are not directly affected by low-level sp$^3$ functionalization, the impact of additional shallow trap states on the temperature dependence is expected to be low. This notion is confirmed by the nearly identical slopes of the logarithmic mobilities *vs.* $1/T$ at temperatures above 150 K. However, for both hole and electron mobilities, the temperature dependence becomes stronger with defect density in the low-temperature range (< 150 K) where charge transport is dominated by tunneling between conductive nanotube segments. As tunneling is independent of temperature, the impact of intra-nanotube transport on the overall network mobility and consequently the impact of the sp$^3$ defects as shallow charge traps should be significant in this temperature range as clearly shown in **Figure 6b**. These data further support the hypothesis that both inter- and intra-nanotube contributions play a significant role for the overall network mobility. If the junction resistances were the only limiting factor for charge transport through the networks, no or only minor differences in the temperature dependence would be observed upon sp$^3$ functionalization.

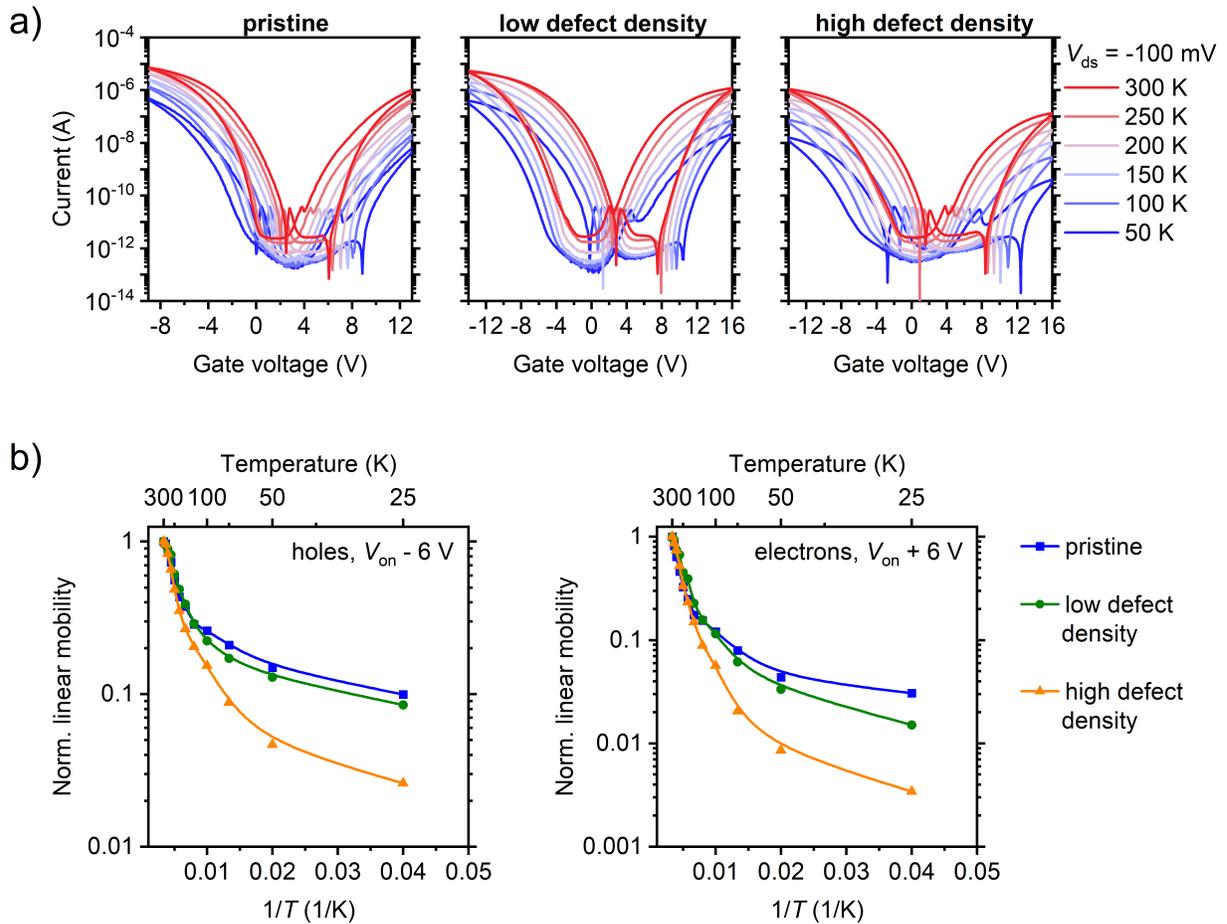

**Figure 6. a)** Temperature-dependent transfer characteristics (four-point probe geometry, $L$ = 40 µm, $W$ = 1 mm, source-drain voltage $V_{ds}$ = -100 mV) of pristine and sp$^3$-functionalized SWCNT network FETs between 25 – 300 K (only every other curve is shown). **b)** Temperature-dependent charge carrier mobilities for holes and electrons normalized to the values at 300 K. For better comparison, all mobilities were contact resistance-corrected and extracted at a fixed gate voltage overdrive of ±6 V for electrons and holes, respectively. Lines are guides to the eye.

The absolute mobility values also decrease with increasing sp$^3$ defect density as shown in **Figure 2e** and Supporting Information, **Figure S23**. Since EL and CMPL measurements indicate that functionalized SWCNT segments are sampled by mobile carriers and contribute to charge transport within the networks, we associate this decrease in mobility directly with increased resistance along individual SWCNTs due to defects rather than a reduction in the number of available current pathways in a dense nanotube network. Electrical measurements on a large number of devices with individual functionalized (6,5) nanotubes (similar to the experiments by Bouilly *et al.*[32]) or short-channel transistors with aligned arrays of functionalized SWCNTs could remove the question of junction resistance. Furthermore, complementary methods such as microwave conductivity measurements[67] or THz spectroscopy[68] would be highly suitable for a contact-free and quantitative determination of intrinsic charge carrier mobilities. A combination of these techniques will be required to unambiguously separate the contributions of sp$^3$ defects and SWCNT junctions to charge transport in functionalized SWCNT networks.

CONCLUSION

We have demonstrated ambipolar and light-emitting field-effect transistors based on networks of polymer-sorted (6,5) SWCNTs with different densities of luminescent sp$^3$ defects that were introduced by diazonium chemistry and led to red-shifted $E_{11}*$ and $E_{11}*^-$ emission. The covalent functionalization of the nanotubes resulted in a moderate, yet gradual decrease of hole and electron mobilities with increasing defect density, thus providing evidence that sp$^3$ defects may act as shallow charge traps that have a noticeable effect on charge transport especially at low temperatures (< 150 K). At room temperature, however, charges can be detrapped fast, and charge modulation PL spectroscopy corroborated that functionalized SWCNT segments are sampled by mobile carriers and actively participate in charge transport through the network. Importantly, even at high defect densities, the FETs remained fully operational and exhibited EL from the defect states *via* ambipolar carrier recombination. The contribution of sp$^3$ defect emission (mostly E$_{11}$*) to electroluminescence was tunable through the defect concentration and stable over a wide range of current densities. As the emission spectrum of the sp$^3$-functionalized SWCNT network should be very narrow for electrically driven light-emitting devices, the selective introduction of only E$_{11}$*$^-$ defects would be highly beneficial for this purpose.[25] Using these defects with a different binding configuration and a deeper exciton trapping potential could also shift the EL spectrum even further toward application-relevant telecommunication wavelengths. Overall, the successful integration of sp$^3$-functionalized semiconducting SWCNT networks into electroluminescent devices may enable applications that require both high carrier mobilities and controllable emission properties, such as electrically pumped single-photon sources in the nIR.

METHODS

**Preparation of (6,5) SWCNT Dispersions.** Nearly monochiral (6,5) SWCNT dispersions were obtained from CoMoCAT raw material (CHASM Advanced Materials Inc., Charge No. SG65i-L58, 0.38 g L$^{-1}$) *via* shear force mixing (Silverson L2/Air mixer, 10230 rpm, 72 h) and polymer-wrapping with poly[(9,9-dioctylfluorenyl-2,7-diyl)-*alt*-(6,6'-(2,2'-bipyridine))] (PFO-BPy, American Dye Source, $M_W$ = 40 kg mol$^{-1}$, 0.5 g L$^{-1}$) in toluene as described previously.[7] The resulting dispersion was centrifuged twice for 45 min at 60000*g* (Beckman Coulter Avanti J26XP centrifuge) and filtered through a polytetrafluoroethylene (PTFE) syringe filter (pore size 5 µm) to remove aggregates.

**sp$^3$ Functionalization of (6,5) SWCNTs.** sp$^3$ Functionalization of PFO-BPy-wrapped (6,5) SWCNTs was performed according to the protocol introduced by Berger *et al.*[34] Briefly, SWCNTs (0.72 mg L$^{-1}$, corresponding to an optical density of 0.4 cm$^{-1}$ at the E$_{11}$ transition) were functionalized with 4-bromobenzenediazonium tetrafluoroborate in an 80:20 vol-% toluene/acetonitrile (MeCN) mixture. To increase the solubility of the diazonium salt, a solution of 18-crown-6 in toluene was added to the (6,5) SWCNT dispersion as a phase-transfer agent (concentration in the final reaction mixture, 7.6 mmol L$^{-1}$). After adding an appropriate amount of the diazonium salt in acetonitrile (concentrations in the final reaction mixture ranging between 0.3 mg mL$^{-1}$ and 2 mg mL$^{-1}$), the reaction was allowed to proceed at room temperature and in the dark for ~16 h. Then, the mixture was filtered through a PTFE membrane filter (Merck Millipore JVWP, pore size 0.1 µm) and the filter cake was washed with MeCN and toluene to remove unreacted diazonium salt, reaction byproducts, and excess polymer.

**Device Fabrication.** Bottom contact electrodes (interdigitated electrodes with $L$ = 20 µm, $W$ = 10 mm and four-point probe geometry with $L$ = 40 µm, $W$ = 1 mm) were patterned by photolithography (double-layer LOR5B/S1813 resist, microresist technology GmbH) and electron beam evaporation of chromium (2 nm) and gold (30 nm) on glass substrates (Schott AG, AF32eco, 300 µm thickness). After lift-off in *N*-methyl pyrrolidone (NMP), substrates were cleaned by ultrasonication in acetone and 2-propanol for 10 min each. Filter cakes of pristine and functionalized SWCNTs were redispersed in fresh toluene by bath sonication for 30 min and the resulting dispersions were spin-coated (3×70 µL, 2000 rpm, 30 s) onto the electrodes with annealing steps (100 °C, 2 min) in between. After rinsing with tetrahydrofuran and 2-propanol to remove residual polymer, SWCNT networks were patterned using photolithography as described above, oxygen plasma treatment (Nordson MARCH AP-600/300, 100 W, 120 s), and lift-off in NMP to remove all SWCNTs outside the channel area. The samples were annealed at 150 °C for 30 min in dry nitrogen atmosphere. As hybrid gate dielectric, ~11 nm of poly (methyl methacrylate) (PMMA, Polymer Source, $M_W$ = 315 kg mol$^{-1}$, syndiotactic) were spin-coated (80 µL, 4000 rpm, 60 s) from a solution in *n*-butyl acetate (6 g L$^{-1}$), followed by deposition of ~61 nm hafnium oxide (HfO$_x$) *via* atomic layer deposition (Ultratech Inc., Savannah S100) at a temperature of 100 °C using a tetrakis(dimethylamino)hafnium precursor (Strem Chemicals Inc.) and water as the oxidizing agent. Thermal evaporation of 30 nm silver top gate electrodes through a shadow mask completed the devices.

**Characterization.** Baseline-corrected absorbance spectra of SWCNT dispersions were recorded with a Cary 6000i UV-vis-nIR spectrometer (Varian Inc.). A Renishaw inVia confocal Raman microscope in backscattering configuration equipped with a 50× long working distance objective (Olympus, N.A. 0.5) was used for acquisition of Raman spectra (532 nm excitation wavelength).

Atomic force micrographs were acquired with a Bruker Dimension Icon in the ScanAsyst mode under ambient conditions. All current-voltage measurements were performed in inert atmosphere using an Agilent 4156C semiconductor parameter analyzer. The effective device capacitances were obtained with an impedance analyzer (Solatron Analytical ModuLab XM MTS) with the transistors operated as plate capacitors (source and drain electrodes shorted and grounded) at a frequency of 100 Hz. Temperature-dependent electrical measurements were performed in vacuum ($< 10^{-6}$ mbar) in a cryogenic probe station with a closed cooling cycle (CRX-6.5 K, Lake Shore Cryotronics Inc.). Starting from the lowest temperature, the temperature was increased in steps of 25 K. Before each measurement, a hold time of 20 min ensured thermal equilibration.

**Optical Measurements.** PL spectra of SWCNT dispersions and network devices were acquired on a home-built laser setup. For excitation, the wavelength-filtered output of a picosecond-pulsed supercontinuum laser (Fianium Ltd. WhiteLase SC400, 20 MHz repetition rate) or the beam of a 785 nm laser diode (Alphalas GmbH, operated in continuous wave mode) were focused on the samples with a nIR-optimized 50× objective (Olympus, N.A. 0.65). Scattered laser light was blocked by appropriate longpass filters. Spectra were recorded with an Acton SpectraPro SP2358 spectrometer (grating blaze 1200 nm, 150 lines mm$^{-1}$) and a liquid nitrogen-cooled InGaAs line camera (Princeton Instruments OMA V:1024). For the acquisition of voltage-dependent PL and EL spectra, SWCNT network transistors were electrically connected and voltages were applied with a Keithley 2612A source meter. All spectra were corrected to account for the absorption of the optics in the detection path and wavelength-dependent detection efficiency. EL images of the transistor channels were recorded with a nIR-optimized 50× objective (Olympus, N.A. 0.65) and a thermoelectrically cooled InGaAs camera (Xenics XEVA-CL-TE3) during constant current sweeps (Keysight B1500A semiconductor parameter analyzer) of the devices. Total electroluminescence intensities during constant current sweeps were measured with a calibrated InGaAs photodiode (Thorlabs FGA-21) and external efficiencies were calculated as reported previously,[69] taking into account the EL spectra and wavelength-dependent photodiode sensitivity.

**Charge Modulation Photoluminescence (CMPL) Spectroscopy.** CMPL spectroscopy was performed as previously described.[54] In brief, SWCNT network transistors were mounted on the laser setup as detailed above. The source and drain electrodes were grounded and the gate bias was modulated with a peak-to-peak voltage $V_{pp}$ around an offset voltage $V_{os}$ using a Keysight 33600A waveform generator, and SWCNT networks were simultaneously excited with a 785 nm laser diode (Alphalas GmbH) operated in continuous wave mode. The spectrally resolved emission was detected with an InGaAs photodiode (Thorlabs FGA10) and the signal was pre-amplified (Femto DLPCA-200 transimpedance amplifier) and sent to a lock-in amplifier (Stanford Research Systems SR830) that was phase-locked to the waveform generator signal. The differential photoluminescence signal $\Delta PL$ was obtained after correction of the spectra to account for the absorption of optics in the detection path and the wavelength-dependent detection efficiency.

## ASSOCIATED CONTENT

**Supporting Information**

Raman and absorption spectra of pristine and sp$^3$-functionalized SWCNT dispersions, AFM image of a pristine SWCNT network, additional electrical characterization (transfer and output curves, linear mobilities, frequency-dependent capacitance) of pristine and sp$^3$-functionalized SWCNT network FETs, EL spectra, voltage- and power-dependent PL spectra, estimation of excitation densities in EL and PL measurements, schematic setup for CMPL spectroscopy, additional voltage- and frequency-dependent CMPL spectra, calculation of trap densities from subthreshold swings, schematic device layout and details for gated four-point probe measurements, temperature-dependent carrier mobilities. **PDF**

## AUTHOR INFORMATION

**Corresponding Author**

*E-mail: zaumseil@uni-heidelberg.de

**ORCID**

Nicolas F. Zorn: 0000-0001-9651-5612

Felix J. Berger: 0000-0003-2834-0050

Jana Zaumseil: 0000-0002-2048-217X

**Notes**

The authors declare no competing financial interest.

ACKNOWLEDGMENTS

This project has received funding from the European Research Council (ERC) under the European Union's Horizon 2020 research and innovation programme (Grant agreement no. 817494 "TRIFECTs").

**FOR TABLE OF CONTENTS ONLY**

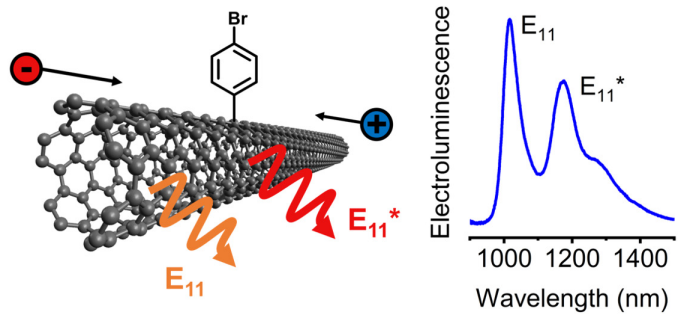

# Supporting Information

# Charge Transport in and Electroluminescence from sp$^3$-Functionalized Carbon Nanotube Networks

*Nicolas F. Zorn, Felix J. Berger, and Jana Zaumseil*[*]

Institute for Physical Chemistry and Centre for Advanced Materials, Universität Heidelberg,

D-69120 Heidelberg, Germany

Corresponding Author

*E-mail: zaumseil@uni-heidelberg.de

# Raman and Absorption Spectroscopy of (6,5) SWCNTs

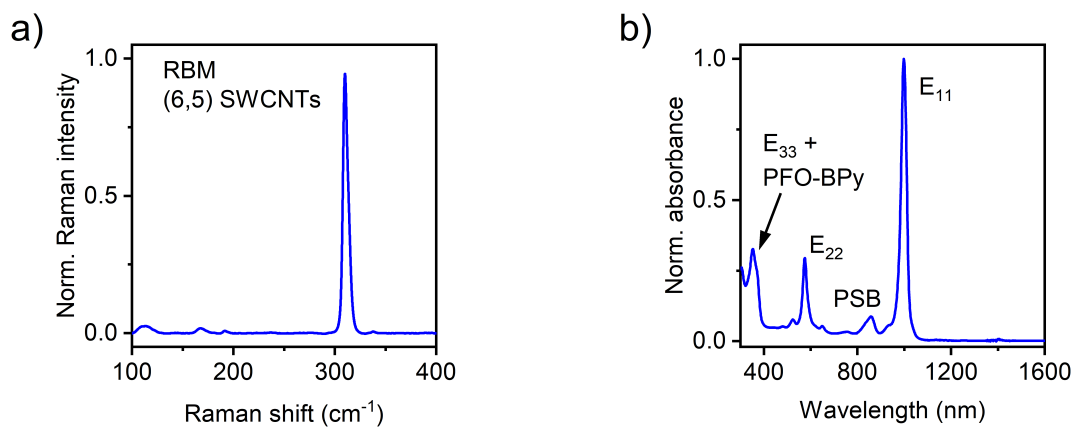

**Figure S1.** Characterization of PFO-BPy-wrapped (6,5) SWCNTs. **a)** Raman spectrum of a drop-cast film of (6,5) SWCNTs in the radial breathing mode (RBM) region. The absence of other peaks confirms the high purity of the dispersion without residual metallic nanotubes or other minority species. The excitation wavelength was 532 nm, and the data was baseline-corrected and averaged over 4000 spectra. **b)** UV-vis-nIR absorption spectrum of a (6,5) SWCNT dispersion in toluene. The main excitonic transitions ($E_{11}$, $E_{22}$, $E_{33}$), the $E_{11}$ phonon side band (PSB), and the absorption band of the wrapping polymer are indicated. No other absorption peaks corresponding to larger-diameter chiralities are observed in the nIR region up to 1600 nm.

# Raman and Absorption Spectroscopy of sp³-Functionalized (6,5) SWCNTs

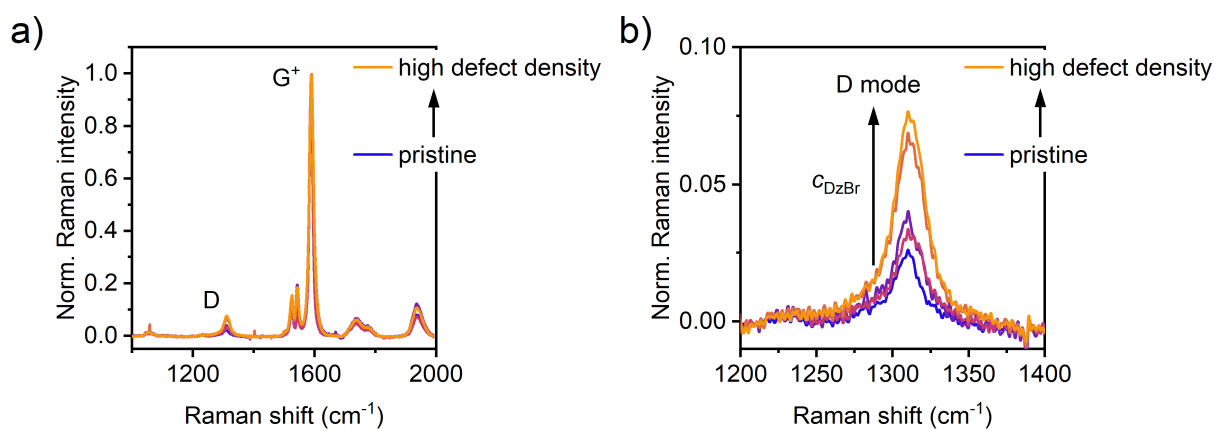

**Figure S2. a)** Raman spectra of drop-cast SWCNT films in the G-mode region and **b)** zoom-in on the D-mode. The increase in the D/G$^+$ area ratio is attributed to a higher density of introduced sp$^3$ defects for higher diazonium salt concentrations ($c_{DzBr}$). For all spectra, the excitation wavelength was 532 nm, and the data was baseline-corrected and averaged over 4000 spectra.

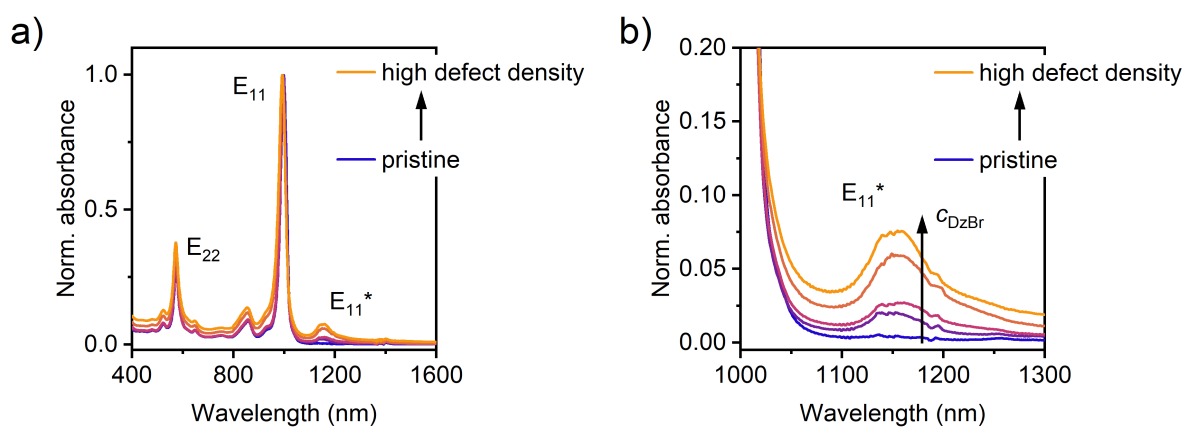

**Figure S3. a)** UV-vis-nIR absorption spectra of pristine and sp$^3$-functionalized SWCNT dispersions. **b)** Zoom-in on the $E_{11}^*$ defect state absorption band that increases with the diazonium salt concentration ($c_{DzBr}$).

**Atomic Force Microscopy**

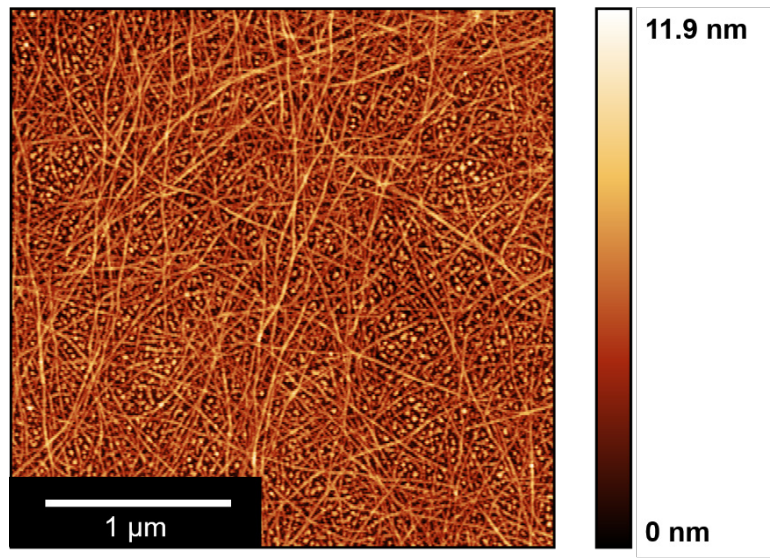

**Figure S4**. Atomic force micrograph of a dense, pristine SWCNT network used for the FET fabrication in this study. Scalebar is 1 µm.

# Electrical Characterization of SWCNT FETs

**Transfer Characteristics and Voltage-Dependent Linear Mobilities**

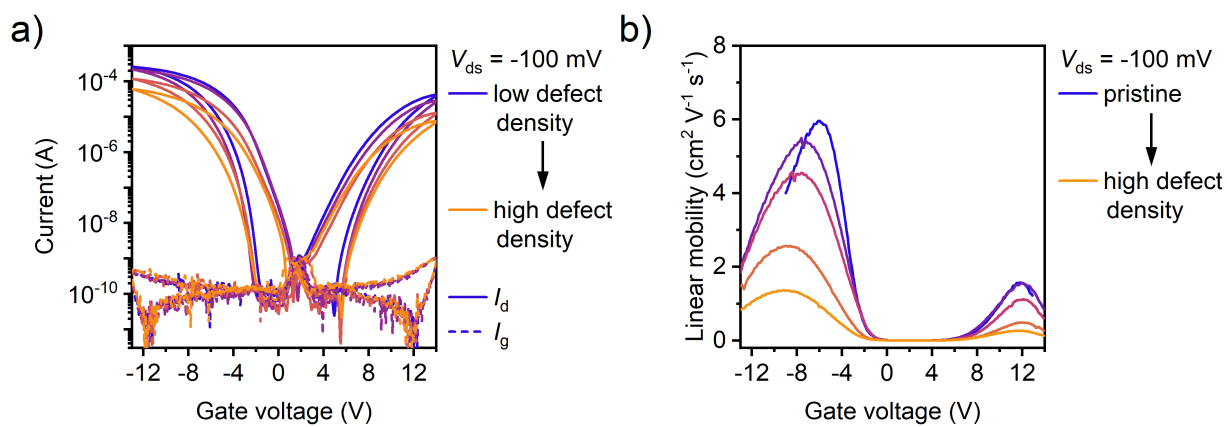

**Figure S5. a)** Representative transfer characteristics (source-drain voltage $V_{ds}$ = -100 mV) of ambipolar FETs based on dense networks of sp$^3$-functionalized (6,5) SWCNTs with different defect densities (drain currents $I_d$, solid lines; gate leakage currents $I_g$, dashed lines). **b)** Voltage-dependent linear charge carrier mobilities of pristine and sp$^3$-functionalized SWCNT network FETs. Only forward sweeps are shown for clarity.

**Output Characteristics**

pristine SWCNT networks

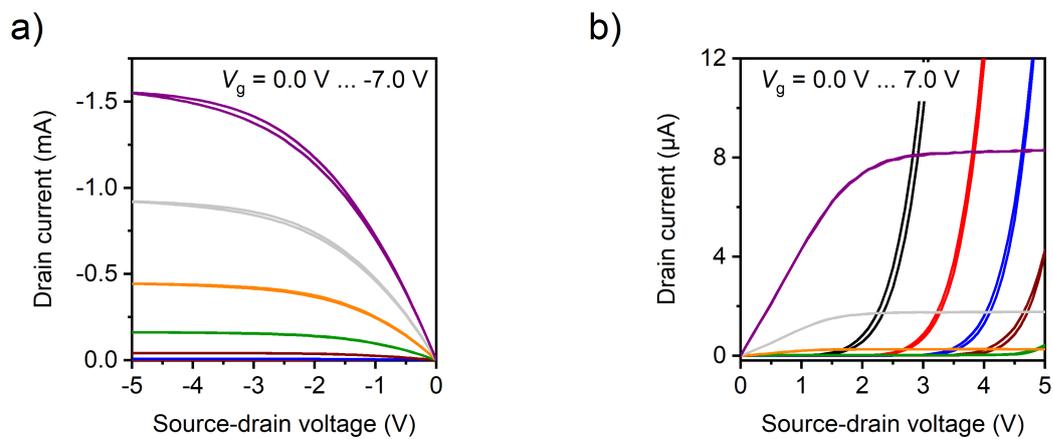

sp³-funct., high defect density

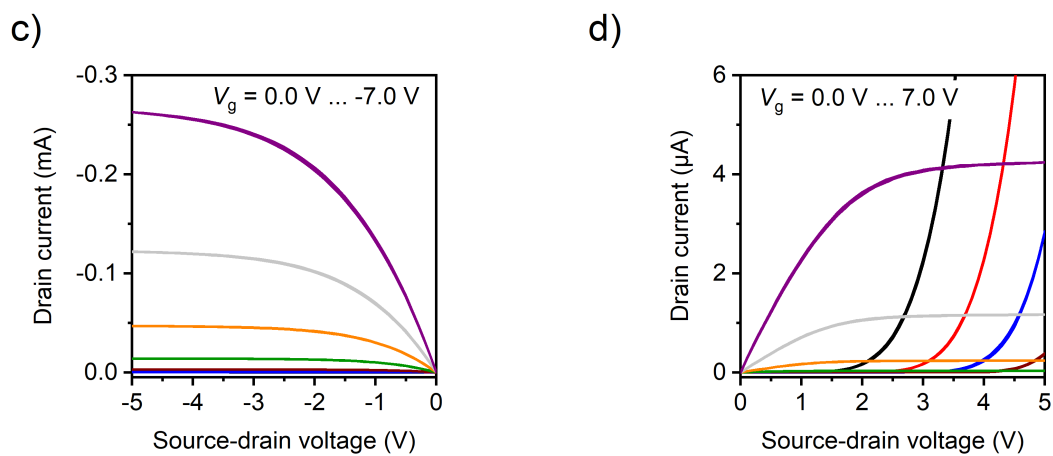

**Figure S6.** Representative output characteristics of ambipolar FETs based on dense networks of **a, b)** pristine SWCNTs and **c, d)** sp³-functionalized SWCNTs with a high defect density in hole and electron accumulation, respectively.

**Linear Mobilities *versus* Defect Metrics**

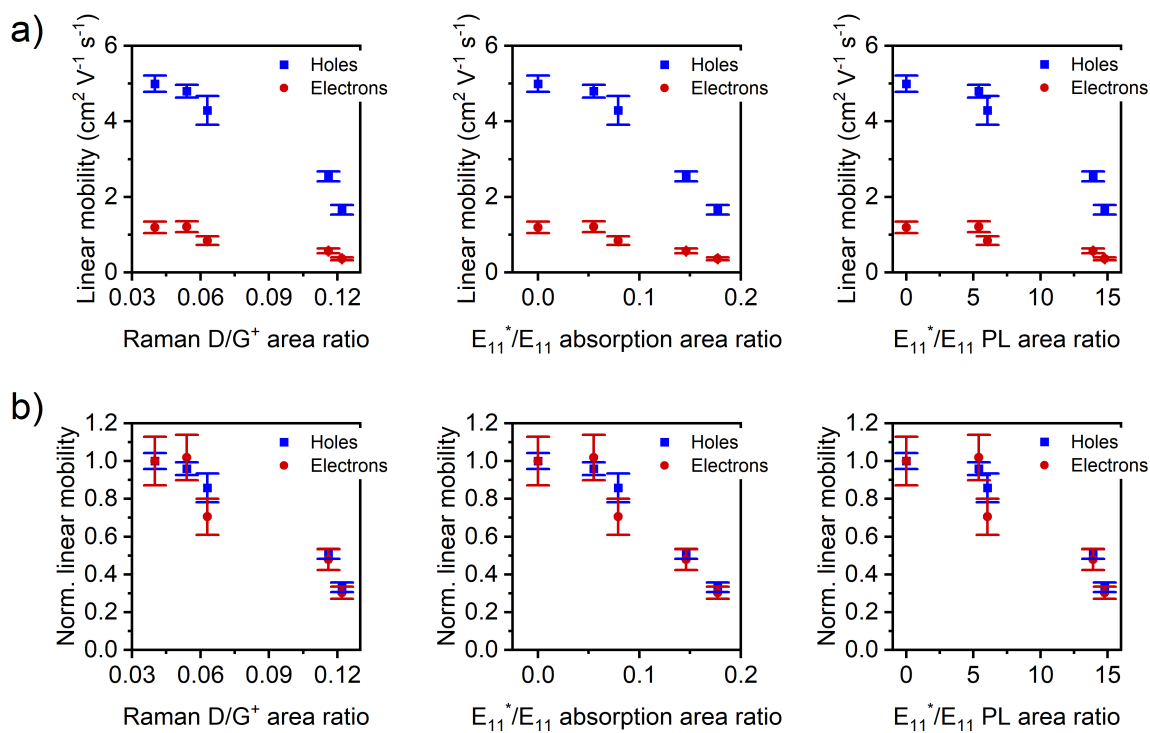

**Figure S7. a)** Absolute linear charge carrier mobilities (holes, blue squares; electrons, red circles) of pristine and sp$^3$-functionalized (6,5) SWCNT network FETs *vs.* the Raman D/G$^+$ area ratio, the $E_{11}^*/E_{11}$ absorption area ratio, and the $E_{11}^*/E_{11}$ PL area ratio, respectively. **b)** Linear charge carrier mobilities normalized to the pristine reference *vs.* defect metrics. Displayed values are maximum carrier mobilities and were averaged over several devices, error bars are standard deviations.

**Frequency-Dependent Capacitance Measurements**

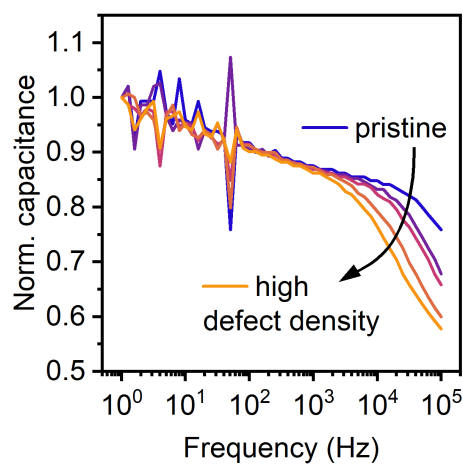

**Figure S8.** Normalized frequency-dependent capacitance of pristine and sp³-functionalized SWCNT network FETs measured in the on-state of the devices (gate voltage $V_g$ = -5 V). The decrease in charge carrier mobilities results in a lower cut-off frequency of the transistors.

**Table S1.** Areal capacitances of pristine and sp³-functionalized SWCNT network FETs extracted in the on-state of the devices.

| Sample | SWCNT network capacitance (nF/cm²) |
|---|---|
| Pristine | 110 |
| Low defect density | 118 |
| Medium-low defect density | 118 |
| Medium-high defect density | 118 |
| High defect density | 117 |

# Spectroscopic Characterization of SWCNT FETs

**Electroluminescence Spectroscopy**

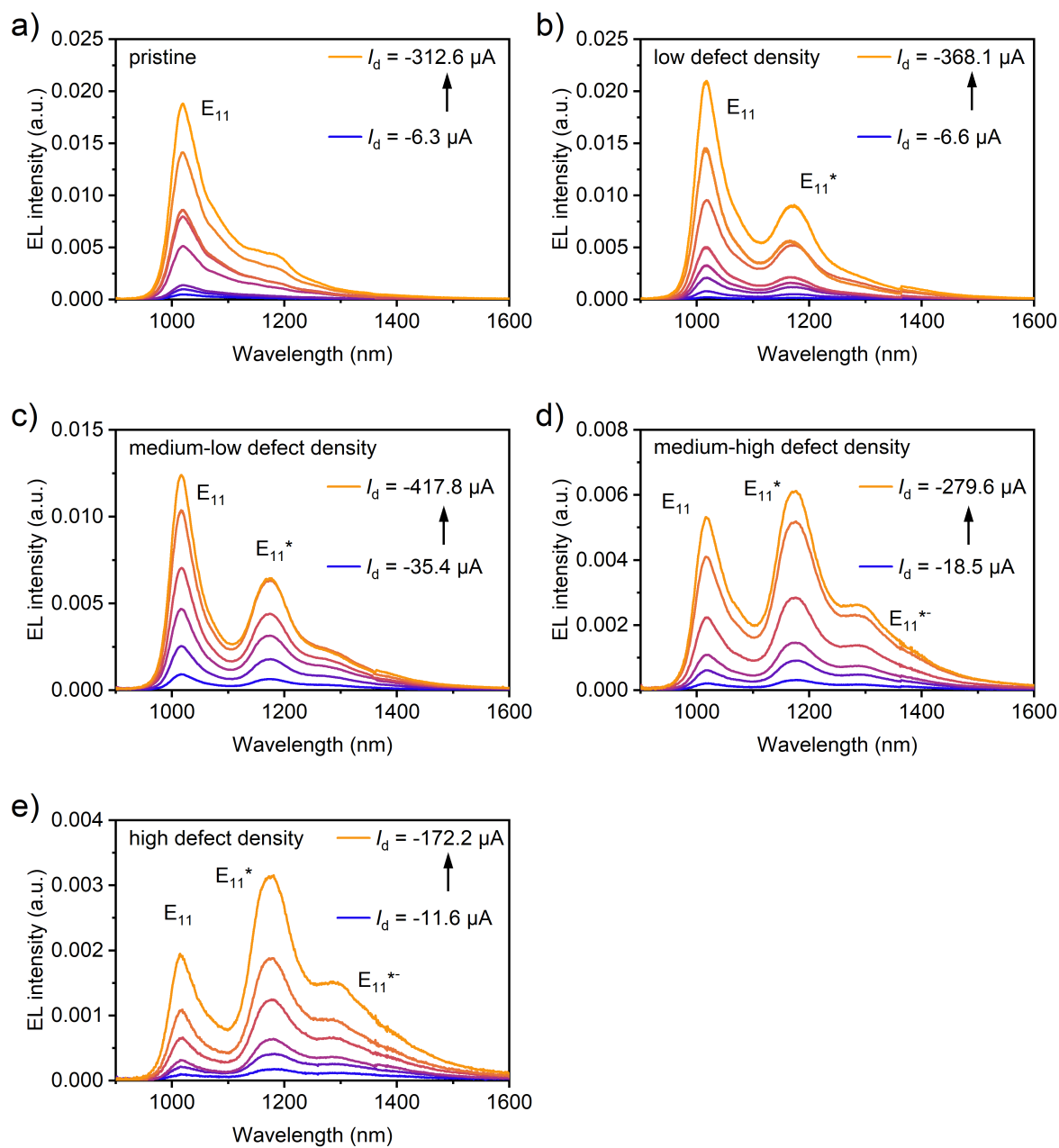

**Figure S9.** EL spectra of **a)** pristine and **b-e)** sp$^3$-functionalized SWCNT network FETs with different defect densities for different drain currents (corresponding gate voltages vary from -2.2 V to -5.0 V). With increasing level of sp$^3$ functionalization, emission from the defect states becomes more dominant.

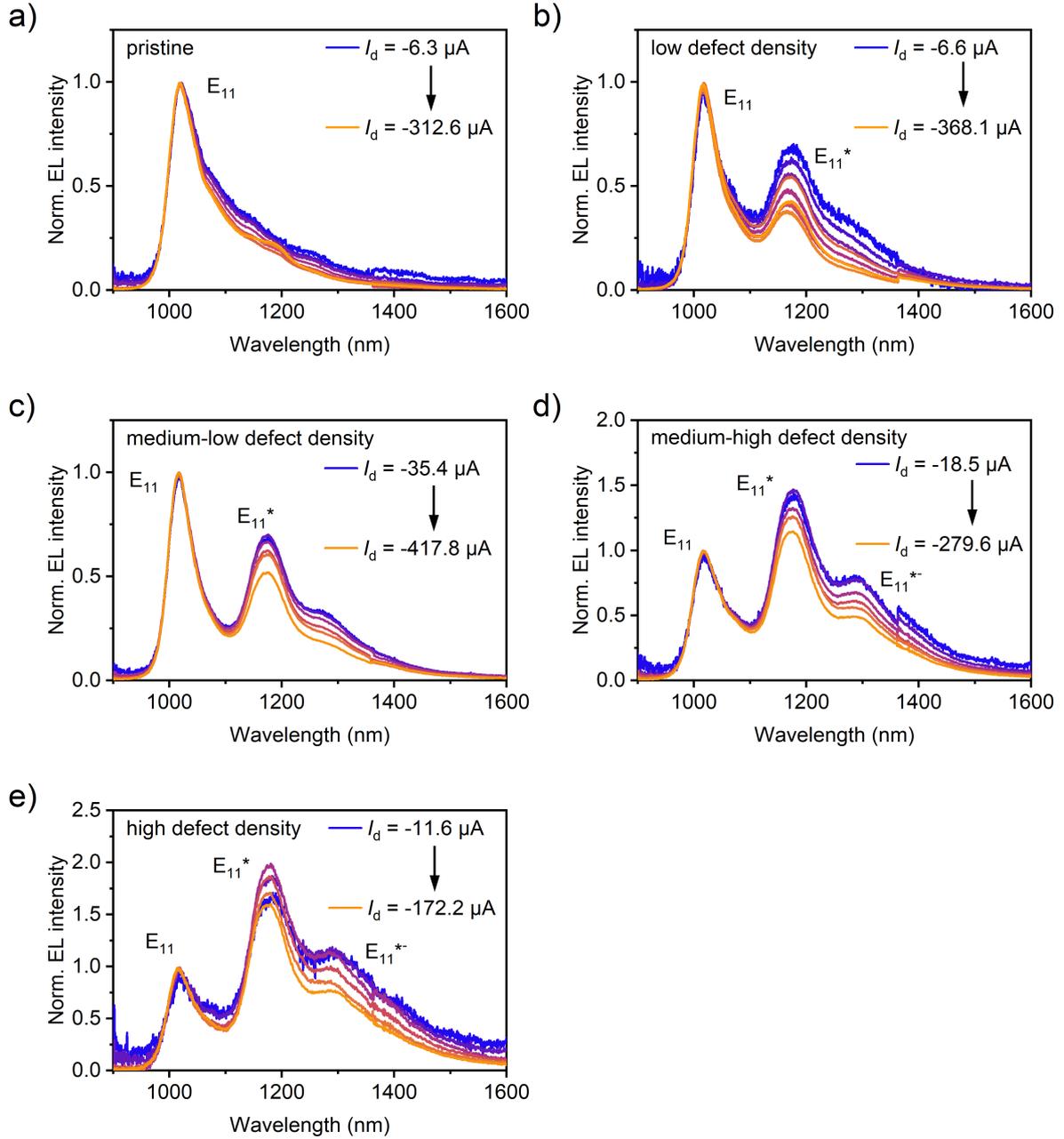

**Figure S10.** EL spectra normalized to the $E_{11}$ exciton peak of **a)** pristine and **b-e)** sp$^3$-functionalized SWCNT network FETs with different defect densities for different drain currents (corresponding gate voltages vary from -2.2 V to -5.0 V). sp$^3$ Defect emission is stable over one to two orders of magnitude in current density. Only a slight decrease of the defect-to-$E_{11}$ emission ratio with increasing drain currents is observed.

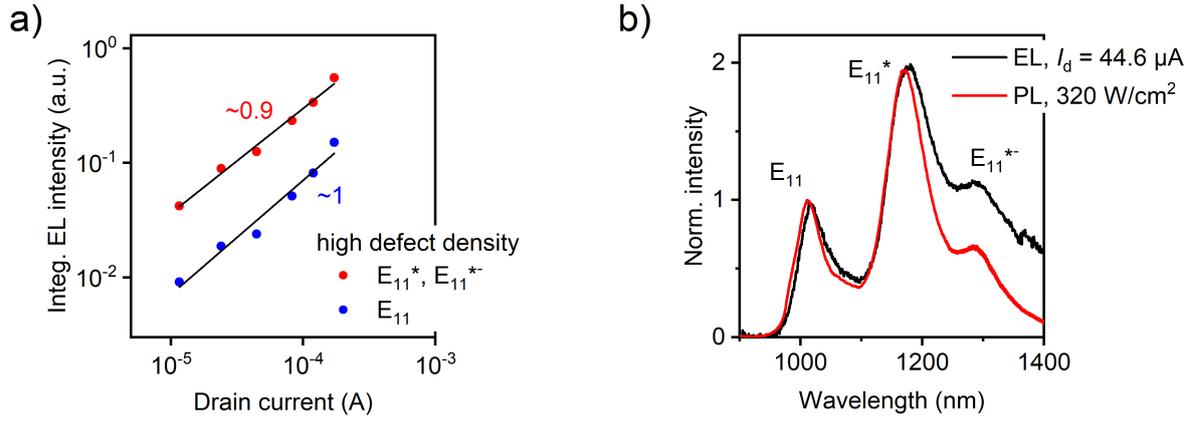

**Figure S11. a)** Log-log plot of the integrated EL intensity of a high defect density sp$^3$-functionalized SWCNT network FET (corresponding spectra shown in **Figure 3e** of the main text) depending on the drain current. The slopes are ~1, indicating that the experiments were conducted in the linear excitation regime where no state-filling is expected. **b)** Comparison between a normalized PL spectrum (low-power, non-resonant continuous wave excitation at 785 nm) and a normalized EL spectrum.

**Voltage- and Excitation Power-Dependent Photoluminescence Spectroscopy**

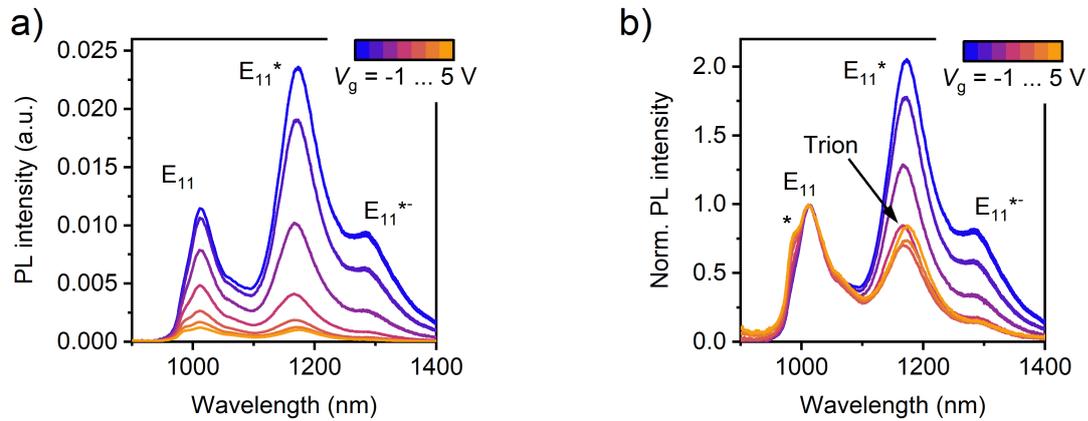

**Figure S12. a)** Static, gate voltage-dependent PL spectra (source-drain voltage $V_{ds}$ = -10 mV) of an sp³-functionalized SWCNT network FET with high defect density in electron accumulation. **b)** Spectra normalized to the $E_{11}$ exciton peak. At high voltages, PL from negatively charged trions can be observed at very similar wavelengths to the $E_{11}^*$ emission. All spectra were acquired under non-resonant continuous wave excitation (785 nm, ~320 W cm⁻²). Note that the peak at ~985 nm marked with an asterisk corresponds to the Raman 2D mode of (6,5) SWCNTs.

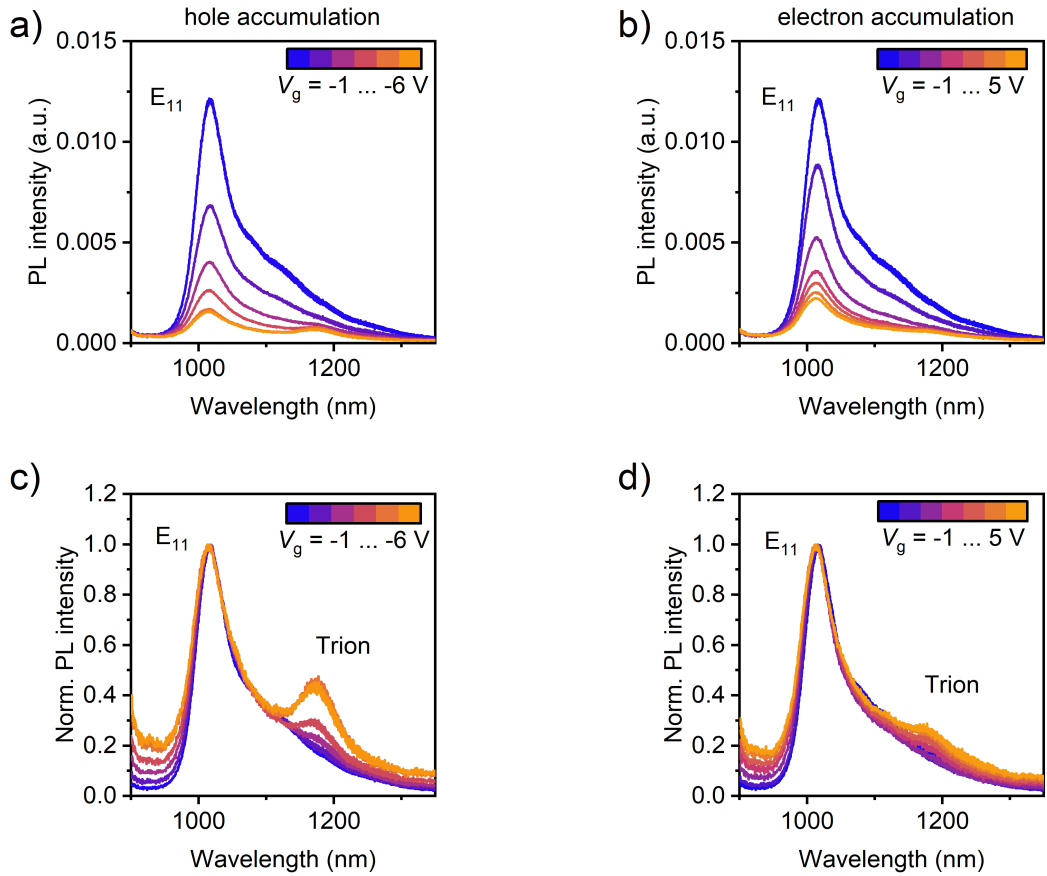

**Figure S13. a, b)** Static, gate voltage-dependent PL spectra (source-drain voltage $V_{ds}$ = -10 mV) of a pristine (6,5) SWCNT network FET in hole and electron accumulation, respectively. **c, d)** Spectra normalized to the $E_{11}$ exciton peak. At high voltages, PL from positively and negatively charged trions, respectively, can be observed. All spectra were acquired under pulsed excitation at the $E_{22}$ transition (575 nm, ~0.02 mJ cm$^{-2}$).

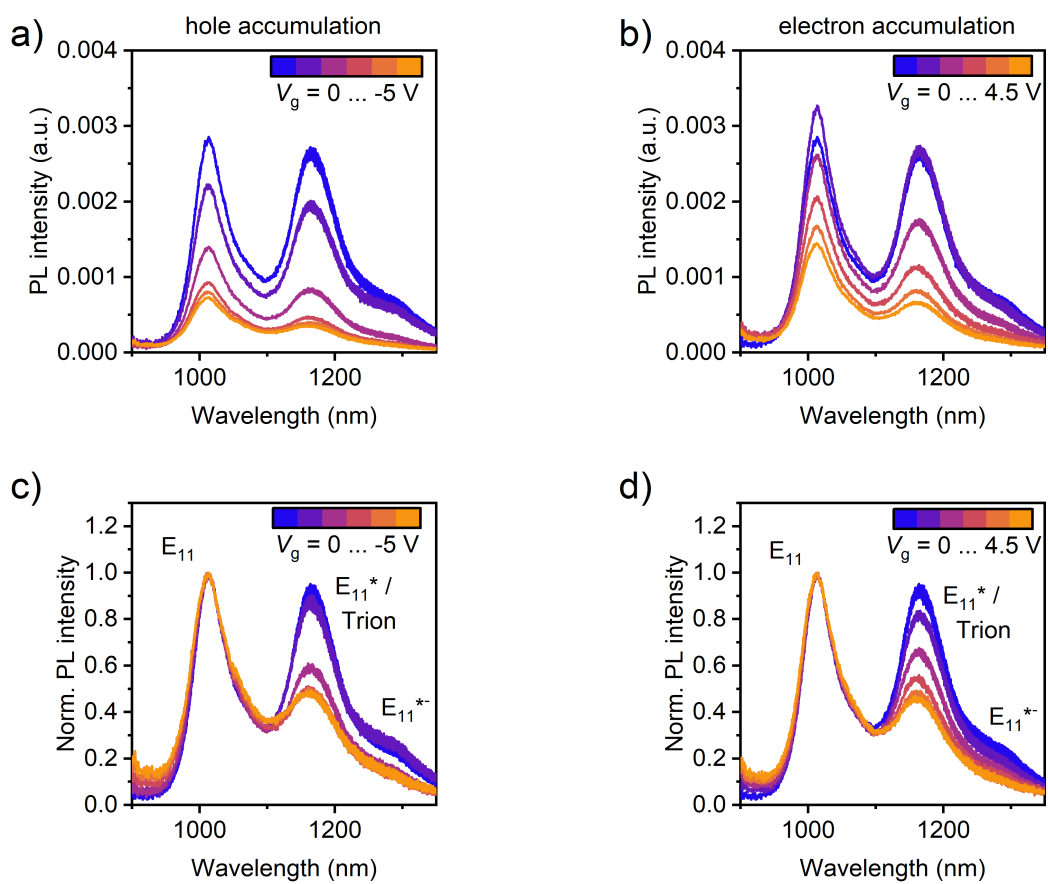

**Figure S14. a, b)** Static, gate voltage-dependent PL spectra (source-drain voltage $V_{ds}$ = -10 mV) of an sp$^3$-functionalized SWCNT network FET with high defect density in hole and electron accumulation, respectively. **c, d)** Spectra normalized to the $E_{11}$ exciton peak. At high voltages, PL from positively and negatively charged trions, respectively, can be observed at very similar wavelengths to the $E_{11}$* emission. All spectra were acquired under pulsed excitation at the $E_{22}$ transition (575 nm, ~0.02 mJ cm$^{-2}$).

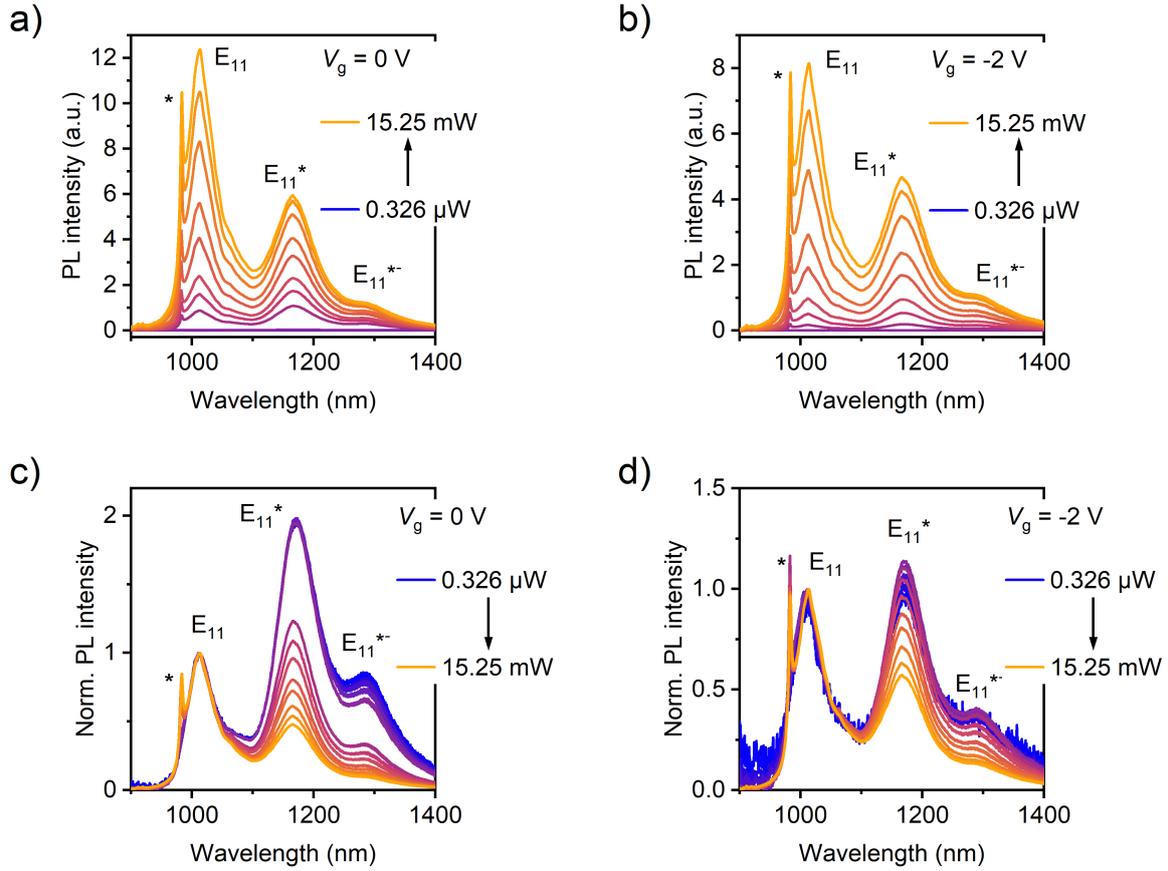

**Figure S15.** Excitation power-dependent PL spectra of an sp$^3$-functionalized SWCNT network FET with high defect density **a)** without applied bias ($V_g$ = 0 V) and **b)** at $V_g$ = -2 V under continuous wave excitation (785 nm). **c, d)** Spectra normalized to the E$_{11}$ exciton peak show saturation of E$_{11}$* and E$_{11}$*$^-$ emission at higher excitation densities. Note that the sharp peak at ~985 nm marked with an asterisk corresponds to the Raman 2D mode of (6,5) SWCNTs.

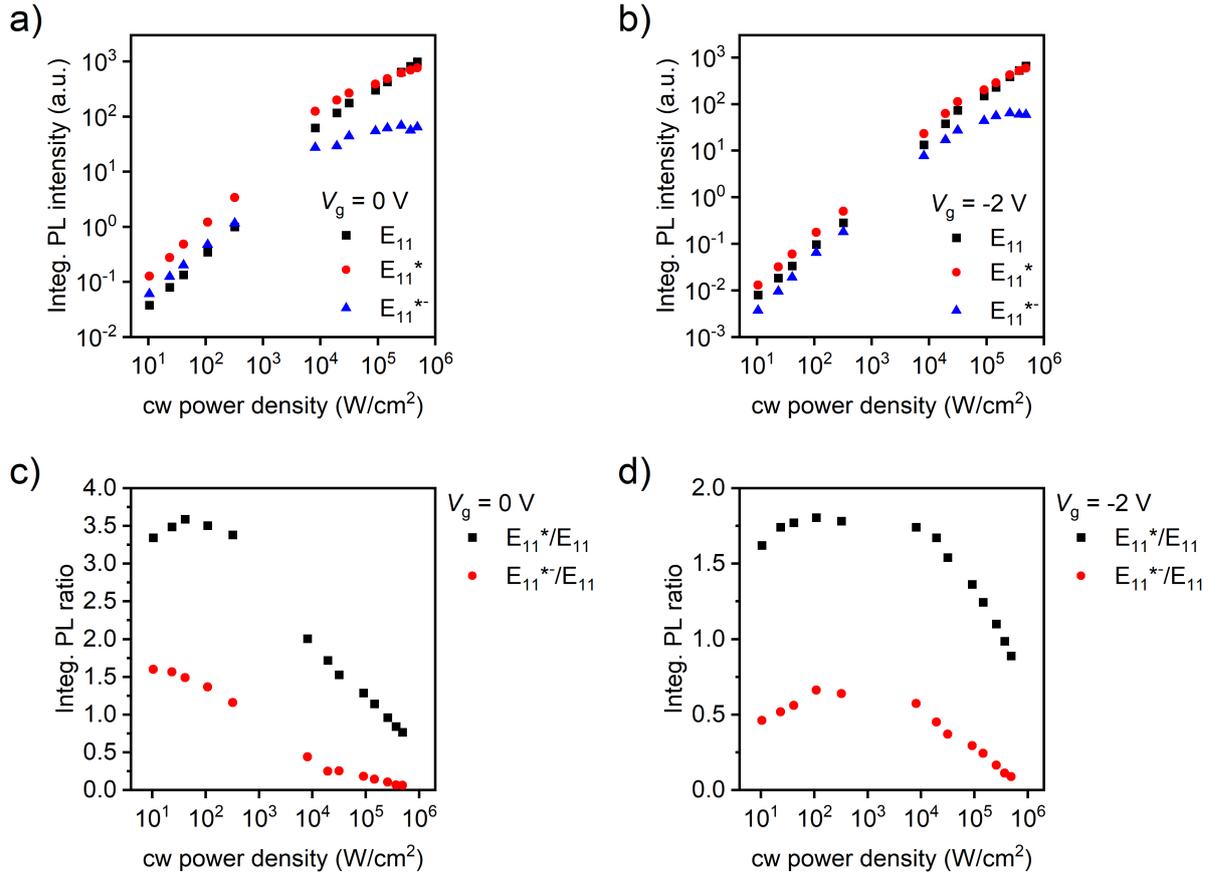

**Figure S16.** Excitation power-dependent, integrated PL intensities of an sp$^3$-functionalized SWCNT network FET with high defect density **a)** without applied bias ($V_g$ = 0 V) and **b)** at $V_g$ = -2 V under continuous wave excitation (785 nm). Intensities were obtained from Lorentzian fits to the spectra in **Figure S15**. **c, d)** Ratios of the integrated PL intensities (E$_{11}$*/E$_{11}$ and E$_{11}$*$^-$/E$_{11}$) show that the emission of E$_{11}$* and E$_{11}$*$^-$ defect states saturates at higher excitation densities when a voltage ($V_g$ = -2 V) is applied **(d)** compared to the neutral state **(c)**.

**Estimation of Excitation Densities in EL and PL Spectroscopy of SWCNT Networks**

To rationalize the differences in defect emission intensities relative to the $E_{11}$ exciton, we estimate the excitation densities in the different experiments as outlined in the following.

In EL measurements, the excitation density is determined by the source-drain current within the transistor channel. A current of 100 μA corresponds to ~$6.2×10^{14}$ charges per second, which is identical to the number of excitons per second assuming that every charge carrier recombines to form an exciton. The total channel width is 10 mm (interdigitated electrodes) and we assume a width of 2 μm for the recombination zone. To calculate the areal density of SWCNTs, we use an average length of ~1.5 μm and a linear density of ~20 μm$^{-1}$ for the dense networks. Thus, the areal density is ~21 SWCNTs per μm$^2$ using the formula from Statz et al.[1] With an exciton lifetime of ~100 ps, we obtain an instantaneous excitation density of ~$1.5×10^{-1}$ excitons per SWCNT. This calculation is identical to Naber et al. except that the quantum yield is not considered here, since we are interested in the overall excitation density rather than the exciton density available for lasing.[2]

In PL experiments with pulsed laser excitation, samples were excited with the output of a supercontinuum laser (20 MHz repetition rate, ~6 ps pulse width, 575 nm excitation wavelength). The average laser power in the excitation spot was ~10 μW, corresponding to ~$1.5×10^6$ photons per pulse. With the photon energy, we can calculate the number of photons per time and per pulse. To estimate the number of absorbed photons, the peak absorption coefficient for excitation at the $E_{22}$ transition as determined by Streit et al.[3] and the geometrical factor of 88,000 carbon atoms per μm nanotube length for (6,5) SWCNTs are used. With an areal density of ~21 SWCNTs per μm$^2$ as detailed above, and a laser spot size of ~2 μm in diameter, we calculate an excitation density of ~$3.4×10^1$ excitons per SWCNT and pulse. This result is in good agreement with Ma et al. who estimated ~3 $E_{11}$ excitons per 1 μW pump laser power for pulsed excitation at the $E_{22}$ transition.[4]

Consequently, the exciton densities under pulsed optical excitation are significantly (~100 times) higher compared to electrical excitation. Due to state filling, sp$^3$ defect emission saturates at lower excitation powers than emission from the mobile $E_{11}$ exciton.[5, 6] Hence, the difference in defect emission intensities (considerably higher defect emission in EL compared to PL under pulsed $E_{22}$ excitation) is inferred to result from the different excitation density regimes.

For PL experiments under continuous wave excitation at 785 nm, the density of generated excitons cannot be estimated since the absorption cross section for this off-resonant excitation wavelength is unknown. However, from the laser power-dependent PL measurements (see **Figures S15 and S16**), it is evident that excitation powers for the acquisition of the spectra in **Figures 4 and S12** were in the linear regime and no state-filling is observed. This notion is further corroborated by the near-identical $E_{11}*/E_{11}$ peak ratios as shown in **Figure S11**, suggesting similar excitation density regimes in EL and non-resonant (785 nm cw excitation) PL experiments.

**Charge Modulation Photoluminescence Spectroscopy**

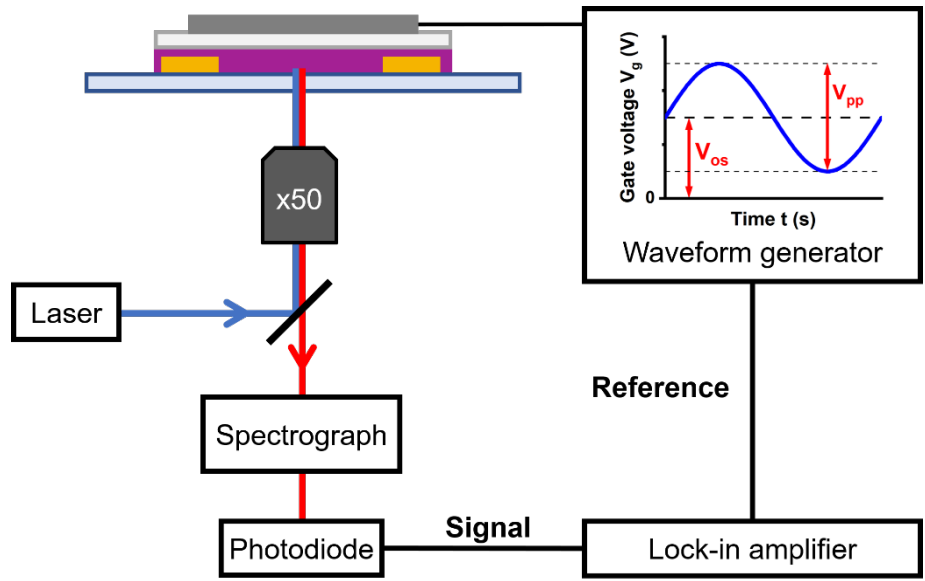

**Figure S17.** Schematic setup for charge modulation photoluminescence (CMPL) spectroscopy. SWCNT network FETs are optically excited with a 785 nm laser diode operated in continuous wave mode. The charge carrier density in the FET channel is modulated with a waveform generator by applying a sinusoidal bias (offset voltage, $V_{os}$; peak-to-peak voltage, $V_{pp}$) to the gate electrode. The emission is spectrally resolved and the signal is detected with an InGaAs photodiode. A lock-in amplifier, which is fed with the sinusoidal bias from the waveform generator, recovers the differential change in PL ($\Delta PL$) by phase-locking the photodiode signal to the reference signal.

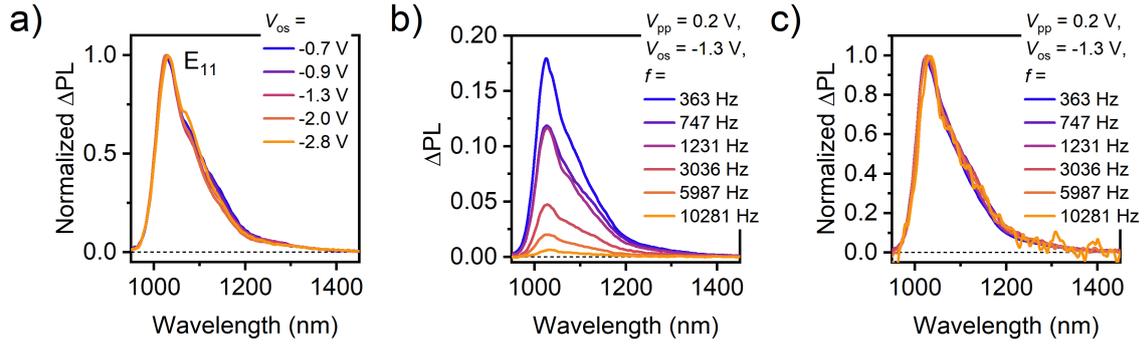

**Figure S18. a)** Normalized voltage-dependent CMPL spectra (absolute intensities shown in **Figure 5a** of the main text) of a pristine (6,5) SWCNT network (modulation frequency $f$ = 363 Hz, peak-to-peak voltage $V_{pp}$ = 0.2 V). **b)** Frequency-dependent CMPL spectra of a pristine SWCNT network and **c)** spectra normalized to the $E_{11}$ $\Delta PL$ signal. The nearly identical normalized spectra confirm the common physical origin (*i.e.*, quenching by mobile charges) of the peaks.

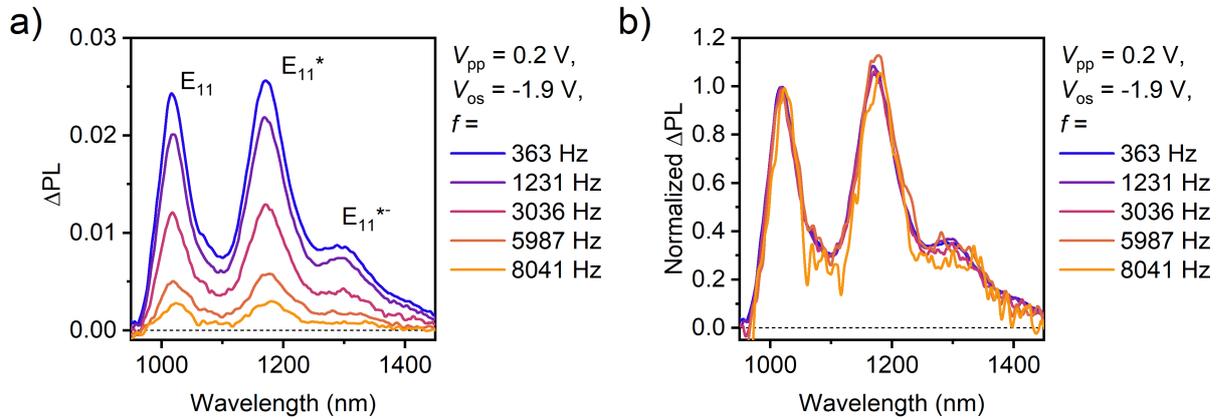

**Figure S19. a)** Frequency-dependent CMPL spectra of an sp$^3$-functionalized SWCNT network with high defect density and **b)** spectra normalized to the $E_{11}$ $\Delta PL$ signal. The nearly identical normalized spectra confirm the common physical origin (*i.e.*, quenching by mobile charges) of the peaks.

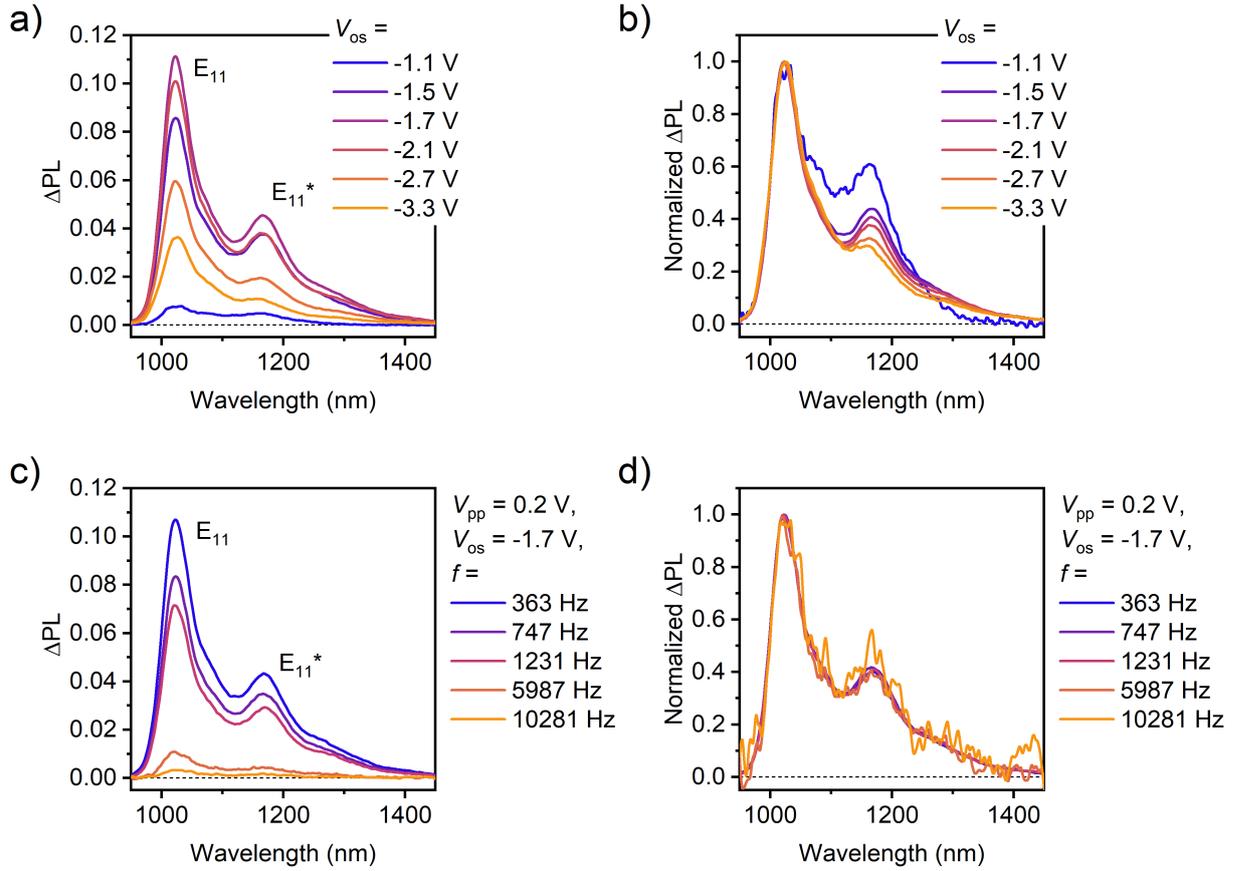

**Figure S20. a)** Voltage-dependent CMPL spectra of a sp$^3$-functionalized SWCNT network with low defect density (modulation frequency $f$ = 363 Hz, peak-to-peak voltage $V_{pp}$ = 0.2 V) and **b)** spectra normalized to the E$_{11}$ $\Delta PL$ signal. **c)** Frequency-dependent CMPL spectra of an sp$^3$-functionalized SWCNT network with low defect density and **d)** spectra normalized to the E$_{11}$ $\Delta PL$ signal. The nearly identical normalized spectra confirm the common physical origin (*i.e.*, quenching by mobile charges) of the peaks.

# Temperature-Dependent Electrical Characterization of SWCNT FETs

## Extraction of Trap Densities from the Subthreshold Regime

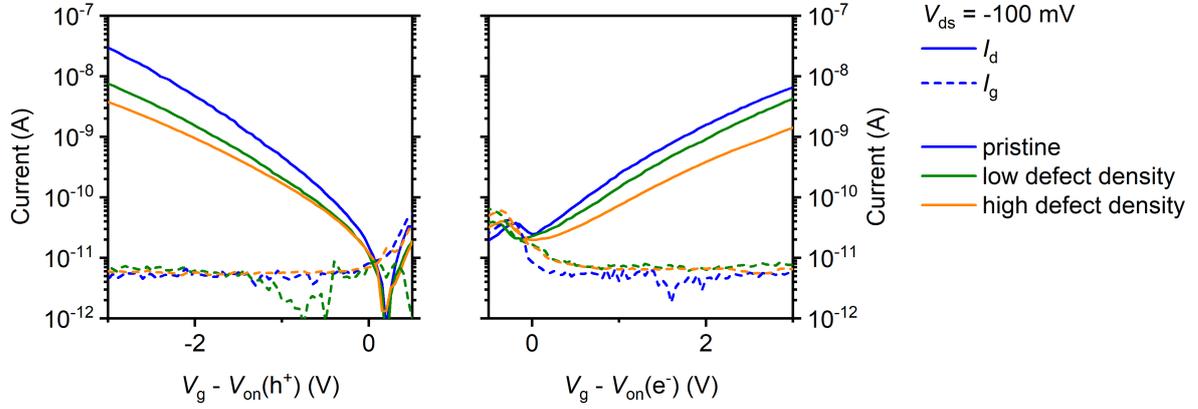

**Figure S21.** Zoom-in on the subthreshold regime in the transfer characteristics (source-drain voltage $V_{ds}$ = -100 mV, $T$ = 300 K) of pristine and sp³-functionalized SWCNT network FETs (drain currents, solid lines; gate leakage currents, dashed lines).

As detailed by Kalb *et al.* the relation between the subthreshold swing $S$ and the trap density $N_\square$ is given by the following formula.[7]

$$S = \frac{k_B T \ln(10)}{e}\left(1 + \frac{e^2}{C_i} N_\square\right) = \frac{\partial V_g}{\partial(\log(I_d))} \tag{1}$$

In Equation (1), $k_B$ is the Boltzmann constant, $T$ is the temperature, $e$ is the elementary charge, and $C_i$ is the areal capacitance. The calculated values for the trap densities are provided in **Table S2**.

**Table S2.** Trap density for holes and electrons calculated from the subthreshold slopes of the transfer characteristics of pristine and sp³-functionalized SWCNT network FETs shown in **Figure S21** according to Equation (1).

|  | Trap density for holes $N_\square(h^+)$ (cm⁻² eV⁻¹) | Trap density for electrons $N_\square(e^-)$ (cm⁻² eV⁻¹) |
|---|---|---|
| Pristine | $(5.9 \pm 0.3) \cdot 10^{12}$ | $(10.1 \pm 0.1) \cdot 10^{12}$ |
| Low defect density | $(8.3 \pm 0.3) \cdot 10^{12}$ | $(13.1 \pm 0.1) \cdot 10^{12}$ |
| High defect density | $(9.4 \pm 0.3) \cdot 10^{12}$ | $(16.0 \pm 0.1) \cdot 10^{12}$ |

**Schematic Device Layout and Principle of Gated Four-Point Probe Measurements**

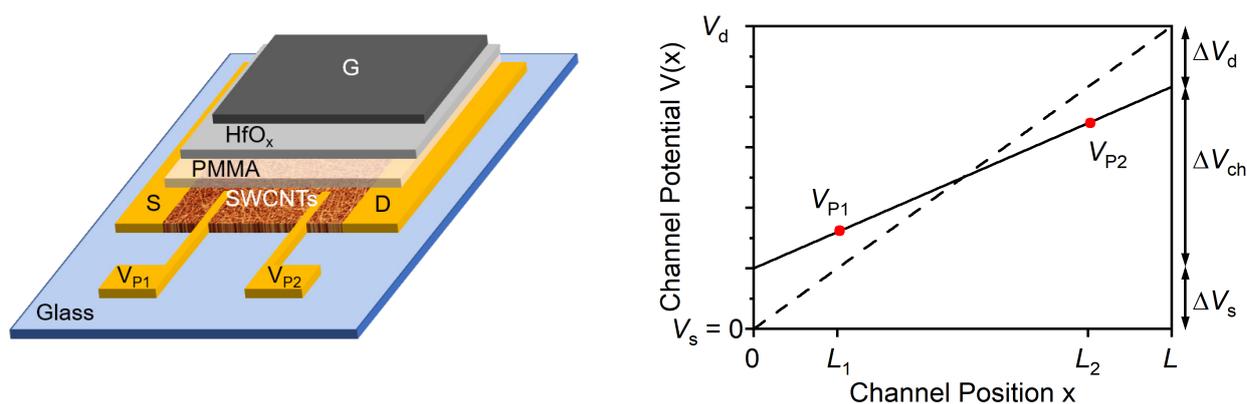

**Figure S22.** Working principle of gated four-point probe measurements. As shown in the schematic device layout (layers are laterally shifted and scaled for better visibility), two voltage probes ($V_{P1}$, $V_{P2}$) are defined at the positions $L_1 = 8$ µm, $L_2 = 32$ µm within the channel ($L = 40$ µm). By linearly extrapolating the potential gradient in the transistor channel measured with the voltage probes (solid line), the potential drops at the source and drain electrodes ($\Delta V_s$, $\Delta V_d$) are determined and thus the contact resistance can be calculated. $\Delta V_{ch}$ is the actual channel potential, and the dashed line corresponds to an ideal case without any contact resistance.

**Temperature-Dependent Mobilities**

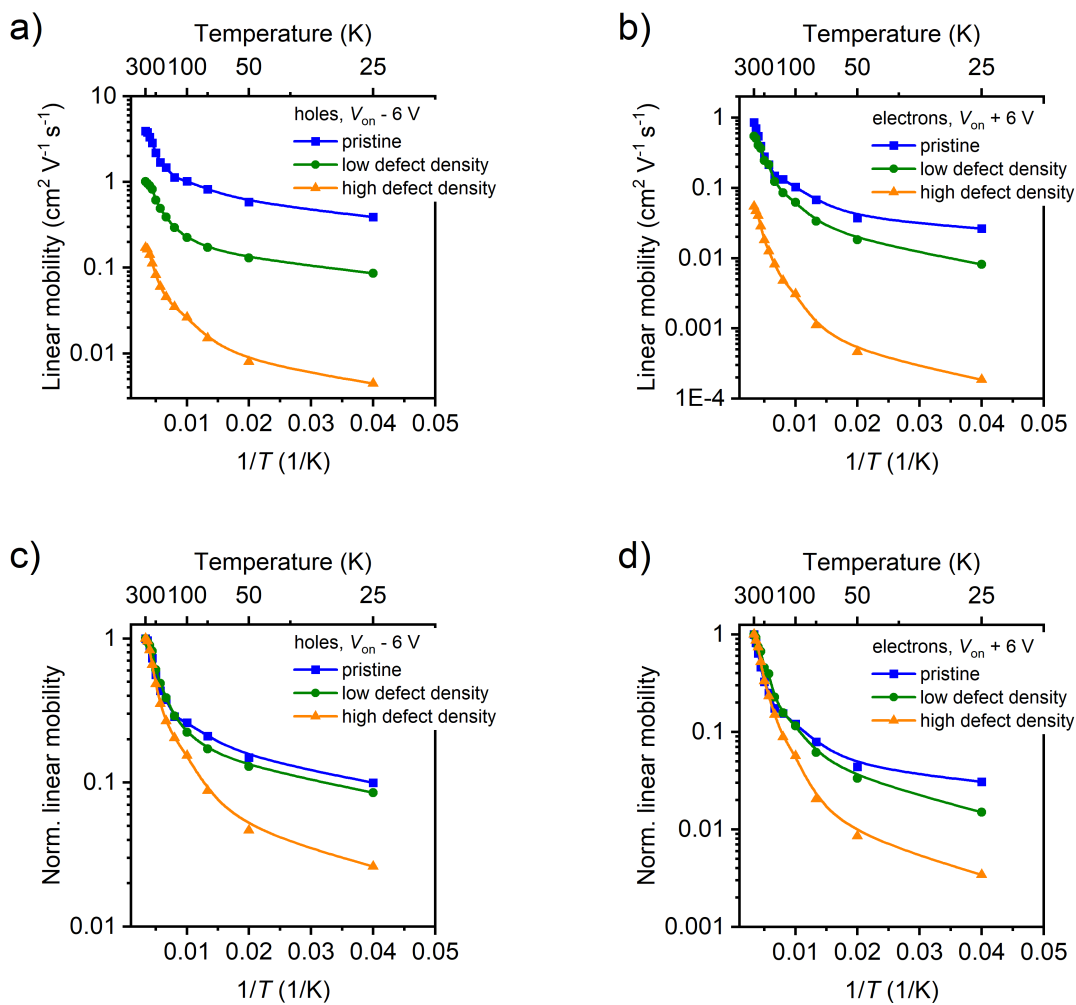

**Figure S23.** Full dataset of temperature-dependent, contact resistance-corrected linear mobilities of pristine and sp$^3$-functionalized SWCNT network FETs. Graphs show **a, b)** absolute and **c, d)** normalized mobilities in the hole and electron transport regimes, respectively. For better comparison, all values were extracted at a fixed gate voltage overdrive of ±6 V for electrons and holes, respectively.